%% file: main.tex
\documentclass{jpp}
\usepackage{graphicx}
\usepackage[utf8]{inputenc}
\usepackage[T1]{fontenc}
\usepackage{amsmath}
\usepackage{color}
\usepackage{overpic}
\usepackage{siunitx}
\usepackage{float}
\usepackage{hyperref}

\newcommand{\figlabel}[1]{\textbf{#1}}
\newcommand{\figlabelw}[1]{\textcolor{white}{\figlabel{#1}}}

\newcommand{\bb}[1]{\boldsymbol{#1}}
\newcommand{\dd}{\mathrm{d}}
\newcommand{\sigmaBK}{\sigma}
\newcommand{\thetap}{\theta}
\newcommand{\Zeff}{Z_{\rm eff}}

\newcommand{\pstar}{{p^\star}}
\newcommand{\tstar}{{\theta^\star}}

\newcommand{\CLISTE}{\textsc{Cliste}}
\newcommand{\CODE}{\textsc{Code}}
\newcommand{\GO}{\textsc{Go}}
\newcommand{\SOFT}{\textsc{Soft}}

\newcommand{\aur}{\mathsf{a}}
\newcommand{\bhat}{\hat{\bb{b}}}
\newcommand{\nhat}{\hat{\bb{n}}}

\newcommand{\perphat}{\hat{\bb{\perp}}}

\newcommand{\ppar}{p_\parallel}
\newcommand{\pperp}{p_\perp}
\newcommand{\rcp}{r_{\text{cp}}}

\newcommand{\rhohat}{\hat{\bb{\rho}}}

\input{authors}
\shorttitle{Runaway current from synchrotron images}
\title{Spatiotemporal analysis of the runaway distribution function from synchrotron images in an ASDEX Upgrade disruption}

\begin{document}

    \maketitle
    
    \begin{abstract}
    Synchrotron radiation images from runaway electrons (REs) in an ASDEX Upgrade discharge disrupted by argon injection are analyzed using the synchrotron diagnostic tool \SOFT\ and coupled fluid-kinetic simulations. We show that the evolution of the runaway distribution is well described by an initial hot-tail seed population, which is accelerated to energies between 25-$\SI{50}{MeV}$ during the current quench, together with an avalanche runaway tail which has an exponentially decreasing energy spectrum. We find that, although the avalanche component carries the vast majority of the current, it is the high-energy seed remnant that dominates synchrotron emission. With insights from the fluid-kinetic simulations, an analytic model for the evolution of the runaway seed component is developed and used to reconstruct the radial density profile of the RE beam. The analysis shows that the observed change of the synchrotron pattern from circular to crescent shape is caused by a rapid redistribution of the radial profile of the runaway density.
    \end{abstract}
    
    \input{section1}
    \input{section2}

    \input{section3}
    \input{section4}
    \input{section5}

    \section*{Acknowledgements}
    The authors are grateful to R.~A.~Tinguely, K.~Insulander Björk, O.~Vallhagen, M.~Dunne, V.~G.~Igochine and M. Maraschek for fruitful discussions, and to M.~Gruber for operating the synchrotron camera diagnostics. This work has been carried out within the framework of the EUROfusion Consortium and has received funding from the Euratom research and training programme 2014 - 2018 and 2019 - 2020 under grant agreement No 633053. This project has received funding from the European Research Council (ERC) under the European Union’s Horizon 2020 research and innovation programme under grant agreement No 647121. The views and opinions expressed herein do not necessarily reflect those of the European Commission. The work was also supported by the Swedish Research Council (Dnr. 2018-03911) and the EUROfusion -- Theory and Advanced Simulation Coordination (E-TASC). G. Pokol acknowledges the support of the National Research, Development and Innovation Office (NKFIH) Grant FK132134.
    
    \appendix
    
    \input{appendixA}
    
    \input{appendixB}

    \bibliographystyle{jpp}
    
    \bibliography{ref}

\end{document}

%% file: authors.tex
\shortauthor{M. Hoppe and others}

\author{%
    M.~Hoppe\aff{1}\corresp{\email{hoppe@chalmers.se}},
    L.~Hesslow\aff{1},
    O.~Embreus\aff{1},
    L.~Unnerfelt\aff{1},
    G.~Papp\aff{2},
    I.~Pusztai\aff{1},
    T.~F\"ul\"op\aff{1},
    O.~Lexell\aff{1},
    T.~Lunt\aff{2},
    E.~Macusova\aff{3},
    P.~J.~McCarthy\aff{4},
    G.~Pautasso\aff{2},
    G.~I.~Pokol\aff{5},
    G.~Por\aff{5},
    P.~Svensson\aff{1},
    the ASDEX Upgrade team\aff{2}\aff{*}
    and the EUROfusion MST1 team\aff{**}
}

\affiliation{%
    \aff{1}Department of Physics, Chalmers University of Technology, Gothenburg, SE-41296, Sweden
    \aff{2}Max Planck Institute for Plasma Physics, D-85748 Garching, Germany
    \aff{3}Institute of Plasma Physics of the CAS, Prague, CZ-18200, Czech Republic
    \aff{4}Physics Department, University College Cork (UCC), Cork, Ireland
    \aff{5}NTI, Budapest University of Technology and Economics, Budapest, HU-1111, Hungary
    \aff{*}see the author list of ``\emph{H.~Meyer~et~al.~2019 Nucl.~Fusion {\bf 59} 112014}''
    \aff{**}see the author list of ``\emph{B.~Labit~et~al.~2019 Nucl.~Fusion {\bf 59} 086020}''
}

%% file: section1.tex
\section{Introduction}\label{sec:introduction}
Understanding runaway electron (RE) dynamics during tokamak disruptions is of utmost importance for the successful operation of future high-current tokamaks, such as ITER \citep{Boozer2015,Lehnen2015,Breizman2019}.   
Disruptions are notoriously hard to diagnose, and existing numerical models cannot simultaneously capture all aspects of their temporally and spatially multiscale nature, including the associated runaway dynamics. Nevertheless, progress towards a reliable predictive capability requires model validation, and therefore finding ways of connecting experimental observations with theoretical predictions is essential.

A powerful and non-intrusive technique for diagnosing relativistic REs in tokamaks is to measure their synchrotron radiation~\citep{Finken1990,Jaspers1993}. The toroidally asymmetric nature of the synchrotron radiation---due to being strongly biased in the direction of motion of the electrons---in addition to its continuum spectrum, help differentiating it from background line radiation using spectral filtering.   
Recent developments of synthetic synchrotron diagnostics have allowed detailed analysis of experimental synchrotron data. Most recently, in \citet{Tinguely2018a,Tinguely2018b,Tinguely2019a}, synchrotron spectra, images and polarisation data were analysed with the help of the synthetic diagnostic \SOFT~\citep{Hoppe2018a} in a series of Alcator C-Mod discharges, providing valuable constraints on runaway energy, pitch angle and radial density. Full-orbit simulations have also recently provided deeper insight into observations of synchrotron radiation in 3D magnetic fields \citep{Carbajal2017,del-Castillo-Negrete2018}.

In this paper, we examine synchrotron emission from REs in discharge \#35628 of the ASDEX Upgrade tokamak, deliberately disrupted using an injection of neutral argon \citep{Pautasso2016}. The injection of argon leads to a rapid cooling---a thermal quench (TQ).  As the TQ duration is shorter than the collision time at the critical velocity for runaway acceleration, a fraction of the most energetic electrons takes too long to thermalize and is left in the runaway region---a mechanism referred to as \emph{hot-tail generation} \citep{Helander2004,Smith2005}. During the subsequent current quench (CQ), 
the initially trace runaway population gets exponentially multiplied through large-angle collisions with the cold thermalized electrons in a \emph{runaway avalanche} \citep{Rosenbluth1997}.  
As the RE beam forms, its synchrotron emission can be observed using fast, wavelength-filtered visible light cameras.  In this particular discharge,  a sudden transition of the synchrotron image from circular to crescent shape was observed during the plateau phase. The probable cause of this spatial redistribution of the current is a magnetic reconnection caused by a (1,1) magnetohydrodynamic (MHD) mode, similarly to the observation by \cite{Lvovskiy2020} on the DIII-D tokamak, using bremsstrahlung X-ray imaging.

We briefly review the relation between the runaway electron distribution function and the observed synchrotron radiation pattern in section~\ref{sec:images}, then present the experimental setup and parameters of the ASDEX Upgrade discharge analyzed in this paper. To determine the spatiotemporal evolution of the runaway electron distribution, we use a coupled fluid-kinetic numerical tool, that takes into account the evolution of the electric field during the CQ self-consistently. This tool, based on coupling the fluid code \GO~\citep{Smith2006,Feher2011,Papp2013} which captures the radial dynamics, and the kinetic solver \CODE~\citep{Landreman2014,Stahl2016} that models the momentum space evolution, is presented in section~\ref{sec:numerics}.   The collision operator used in \CODE\ includes detailed models of partial screening \citep{Hesslow2018generalized}, which is particularly important in this case, due to the presence of a large amount of partially ionized argon. 

Using the electron distribution function obtained by the coupled fluid-kinetic simulation, we show in section~\ref{sec:numerics} that the resulting synchrotron radiation, computed with \SOFT, 
agrees well with the observed image, both regarding its shape, as well as the growth and spatial distribution of the intensity. Furthermore, inspired by the numerical simulations, we develop an analytical model, which is used in section~\ref{sec:currentprofile}, to reconstruct the radial density profile of the RE beam.
The analysis shows that the change of the synchrotron pattern from circular to crescent shape is
caused by a rapid redistribution of the radial profile of the RE density.

%% file: section2.tex
\section{Synchrotron radiation from runaway electrons}\label{sec:images}
When observed using a fast camera, the synchrotron radiation emitted by runaway electrons typically appears as a pattern on one side of the tokamak central column. The size and shape of the synchrotron pattern is directly related to the energy, pitch and position of the runaway electrons---the runaway electron distribution function. 
Disentangling such dependencies has been the subject of several studies~\citep{Pankratov1996,Zhou2014,Hoppe2018a,Hoppe2018b}; here we briefly review the aspects that are most important for our present purposes.

\subsection{Interpretation of the synchrotron pattern}
Although runaway electrons tend to occupy a large region in momentum space, the corresponding synchrotron pattern can often be well characterised with the energy and pitch angle of the particle 
that has the highest contribution to the camera image. 
We refer to 
these
as the \emph{dominant} or \emph{super} particle of the pattern. We therefore define the super particle as the momentum space location 
 $(p,\thetap)$ that maximises the quantity
\begin{equation}\label{eq:Ihat}
    \hat{I} = G(r,p,\thetap) f(r,p,\thetap) p^2\sin\thetap,
\end{equation}
where $f(r,p,\thetap)$ is the runaway electron distribution function and $p^2\sin\thetap$ is the momentum space Jacobian in spherical coordinates. The Green's function $G(r,p,\thetap)$ quantifies the radiation received by a camera from a particle on the orbit labelled by radius $r$, with momentum $p$ and pitch angle $\thetap$ (given at the point along the orbit of weakest magnetic field).

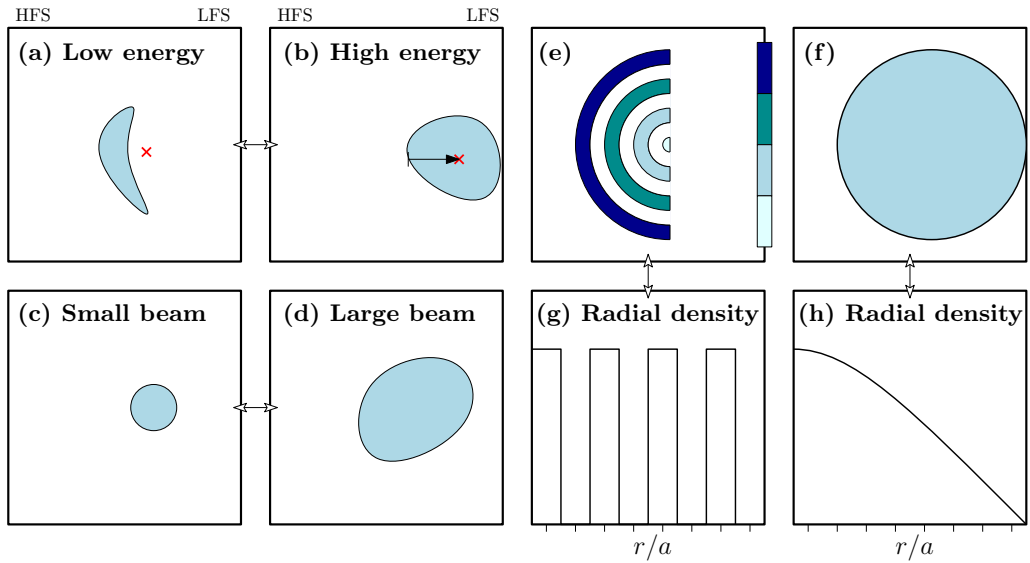
\begin{figure}
    \centering
    \begin{overpic}[width=\textwidth]{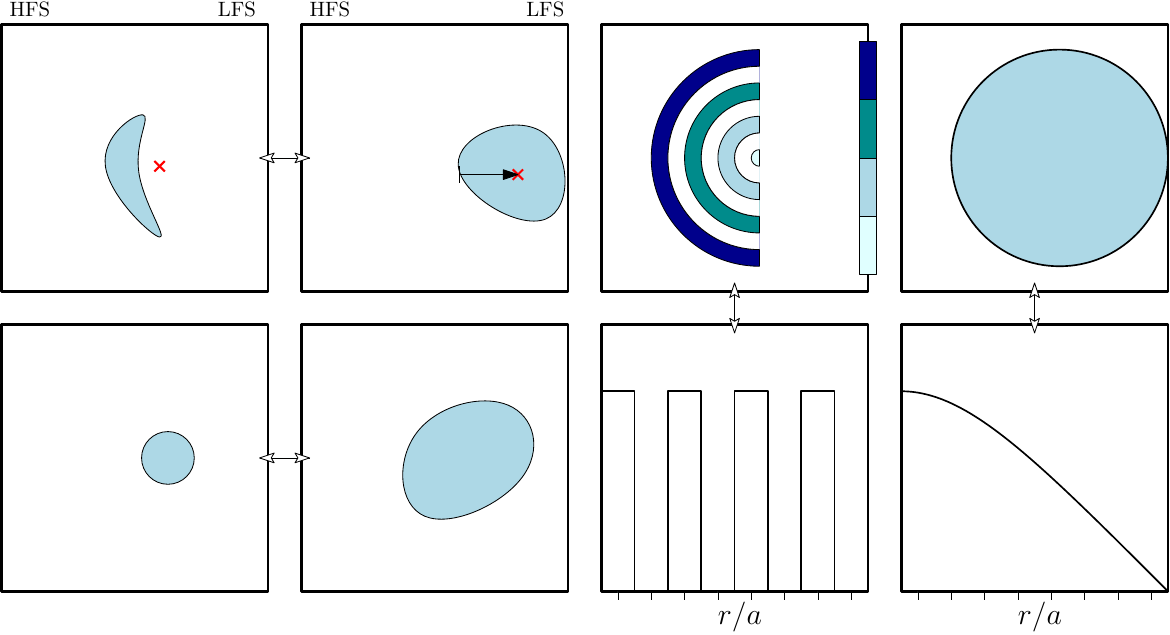}
        \put(1,49){\figlabel{(a) Low energy}}
        \put(27,49){\figlabel{(b) High energy}}
        \put(52,49){\figlabel{(e)}}
        \put(78,49){\figlabel{(f)}}
        \put(1,23.5){\figlabel{(c) Small beam}}
        \put(27,23.5){\figlabel{(d) Large beam}}
        \put(51.8,23.5){\figlabel{(g) Radial density}}
        \put(77.5,23.5){\figlabel{(h) Radial density}}
    \end{overpic}
    \caption{
        (a/b) Illustration of how the runaway energy affects the observed synchrotron pattern.
        At low energies, most radiation originates from the high-field side (HFS), while at higher
        energies, a significant amount of radiation can also be seen on the low-field side (LFS).
        (c/d) The runaway beam radius primarily determines the synchrotron pattern radius.
        (e/g) Illustration of how the runaway radial density affects the synchrotron pattern (darker colours indicate more radiation).
        (f) If the radial density is decreasing with $r/a$, as in (h), synchrotron radiation from low energy
        runaways, such as in (a), could take on a more uniform intensity distribution.
    }
    \label{fig:energy-radius}
\end{figure}

In present-day tokamaks the synchrotron spectra typically peak at infra-red wavelengths, and only a small fraction of the emitted intensity falls into the visible range. The visible-light intensity is, however, usually sufficient to be clearly distinguished from the background radiation, thus it is common to use visible-light cameras for synchrotron radiation imaging.
The \emph{observed} short wavelength tail of the spectrum is exponentially sensitive to the magnetic field strength \citep{Hoppe2018b}, causing the fraction of emitted visible light to sometimes vary by orders of magnitude as an electron travels from the low-field side to the high-field side of the tokamak. As a result, whenever the synchrotron peak of the spectrum is located far from the spectral range of the camera, a crescent-like pattern emerges, as illustrated in figure~\ref{fig:energy-radius}a. 

In a fixed detector/magnetic field setup, this effect can be thought of as an indicator of the runaway energy, as the peak wavelength of the synchrotron spectrum scales as $1/(\gamma^2 B\sin\thetap)$, where $\gamma$ is the Lorentz factor of the electron and $B$ the magnetic field strength.\footnote{While the peak also depends on pitch angle, the pitch angle additionally alters the vertical and toroidal extent of the pattern, thus clearly distinguishing a change in energy from a change in pitch angle.} A sketch of two typical pattern shapes at low and high runaway energy (relative to the camera spectral range)  are shown in figure~\ref{fig:energy-radius}a and b, respectively.

Figures~\ref{fig:energy-radius}a and b also illustrate another important consequence of changing the energy, which is related to the guiding-center drift motion. At higher energies, the guiding-center orbits shift significantly towards the outboard side of the tokamak, as does the corresponding synchrotron pattern. 
Although the guiding-center drift motion is routinely solved for in modern orbit following codes, accurately accounting for the effects of drifts in simulations of synchrotron radiation images is non-trivial and has, to our knowledge, previously only been employed in calculating the effect of synchrotron radiation-reaction~\citep{Hirvijoki2015}. The details of the recently implemented support for guiding-center drifts in \SOFT\ are provided in Appendix~\ref{app:drifts}.

Synchrotron patterns are also sensitive to the spatial distribution of runaway electrons, which will be utilised in Section~\ref{sec:currentprofile}.   In \SOFT, toroidal symmetry is assumed, along with that the poloidal transit time of a runaway is much shorter than the collision time. This leaves the minor radius as a single spatial coordinate for the parametrization of guiding-centre orbits, taken here to be the minor radius $r$ where the electron passes through the outer midplane.

The larger the radius of the runaway beam, the larger the size of the corresponding synchrotron pattern, as illustrated in figures~\ref{fig:energy-radius}c and d. 
In a simulation with a runaway population distributed radially in a series of rectangle functions, as in figure~\ref{fig:energy-radius}g, the synchrotron pattern will be similar to the sketch in figure~\ref{fig:energy-radius}e, where darker colours correspond to higher---and white to zero---observed intensity. Thus, each radial point $r$ contributes a thin band of radiation, weighted by the value of the distribution function in that point. The semi-circular shape  illustrates that, at low runaway energies, the radiation intensity is greater on the high-field side (as in figure~\ref{fig:energy-radius}a) and at larger radii. As a consequence, if the radial density decreases with $r$, as in figure~\ref{fig:energy-radius}h, the corresponding synchrotron pattern can appear to have uniform intensity across all radii, as in figure~\ref{fig:energy-radius}f.

A large variety of synchrotron patterns have been reported in the literature, although circular and crescent patterns seem to be among the more common ones. In this paper, and in particular in section~\ref{sec:currentprofile}, we will analyse the transition from a circular synchrotron pattern into a crescent pattern and find, as suggested above, that the transition is due to a redistribution of the runaway density. Similar transitions have been observed before, most recently by~\cite{Lvovskiy2020} who observe a similar sub-millisecond transition as observed here. Earlier reports also show that transitions from ellipses to crescents, and vice versa, can occur over longer time scales in both disruptions~\citep{Hollmann2013} and quiescent flat-top plasmas~\citep{England2013}.

\subsection{Experimental setup}
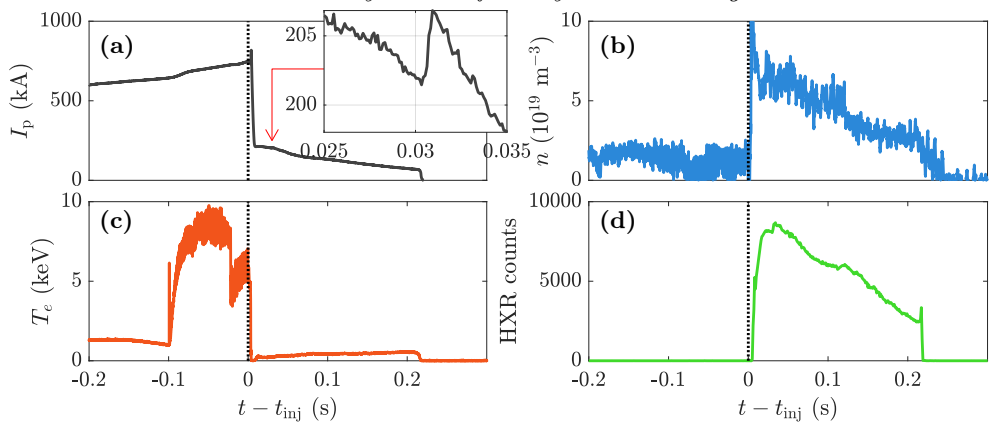
\begin{figure}
    \centering
    \begin{overpic}[width=\textwidth]{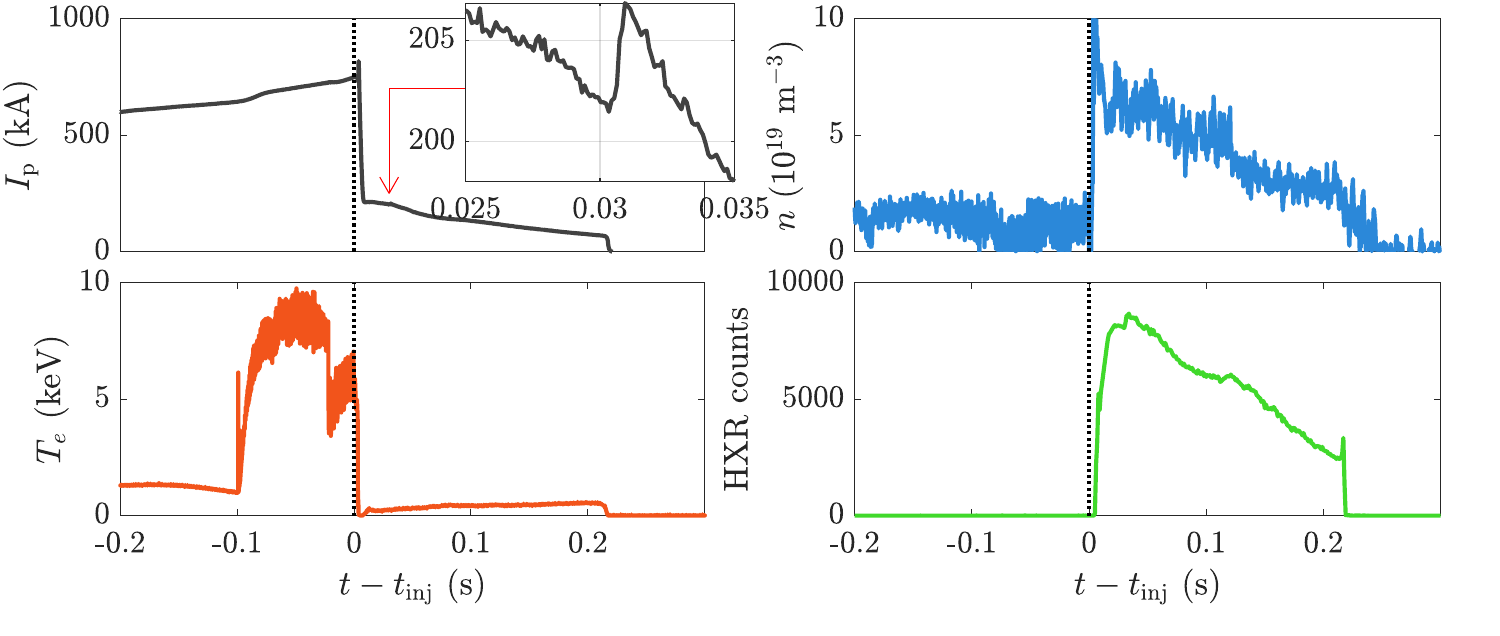}
        \put(9,38){\figlabel{(a)}}
        \put(58,38){\figlabel{(b)}}
        \put(9,21){\figlabel{(c)}}
        \put(58,21){\figlabel{(d)}}
    \end{overpic}
    \caption{Overview of the most relevant plasma parameters in ASDEX Upgrade discharge \#35628. (a) Total plasma current, with the smaller, zoomed-in figure showing a small secondary current spike, (b) line-averaged electron density from central chord CO$_2$ interferometry, (c) electron temperature from central electron cyclotron emission (note that the temperature decreases somewhat just before the disruption to approximately $\SI{4.7}{keV}$), (d) ex-vessel hard x-ray counts.}
    \label{fig:overview35628}
\end{figure}

ASDEX Upgrade is a medium-sized tokamak (major radius $R = \SI{1.65}{m}$, minor radius $a = \SI{0.5}{m}$) located at the Max Planck Institute for Plasma Physics in Garching, Germany \citep{Meyer2019}.
An overview of plasma current, electron density, electron temperature and the hard x-ray count rate in ASDEX Upgrade discharge \#35628 is shown in figure~\ref{fig:overview35628}. This circular, L-mode discharge with $\SI{2.5}{MW}$ ECRH core electron heating applied $\SI{100}{ms}$ before the disruption was deliberately triggered by injecting $N_{\rm Ar}\approx 0.98\times10^{21}$ argon atoms into the plasma at $t=\SI{1}{s}$. Due to the circular plasma shape, the plasma was vertically stable during the discharge, consistent with diagnostic camera recordings. About $\SI{3}{MW}$ of ICRH heating was also applied for $\SI{200}{ms}$ before the disruption, as part of a different experiment, in a configuration where the power was poorly coupled. The discharge developed a subsequent runaway plateau with a starting current of $\approx \SI{200}{kA}$ and a duration of $\approx \SI{200}{ms}$. Before the disruption, the plasma current was $I_{\rm p} \approx \SI{800}{kA}$, the on-axis toroidal magnetic field was $B_{\rm T} = \SI{2.5}{T}$, the central electron temperature was $T_e = \SI{4.7}{keV}$, and the central electron density was $n_e = \SI{2.6e19}{\per\cubic\metre}$. A drop in electron temperature is observed shortly before the disruption due to internal mode activity.

\begin{figure}
    \begin{center}
        \begin{overpic}[width=0.32\textwidth]{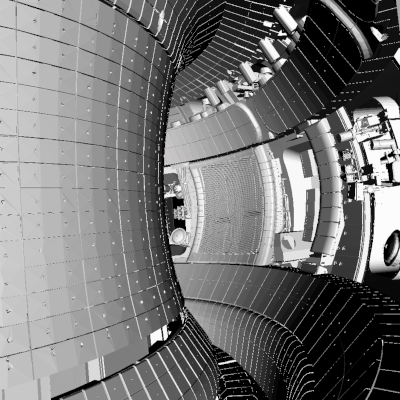}
            \put(5,90){\figlabelw{(a) Camera view}}
        \end{overpic}
        \begin{overpic}[width=0.32\textwidth]{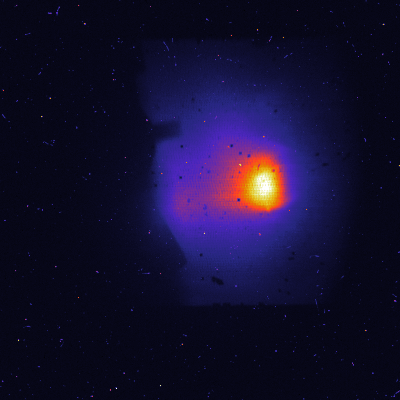}
            \put(5,90){\figlabelw{(b) t = 1.029 s}}
        \end{overpic}
        \begin{overpic}[width=0.32\textwidth]{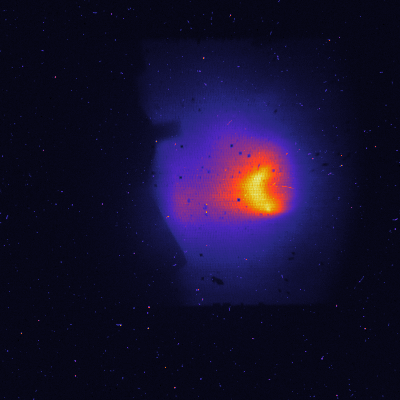}
            \put(5,90){\figlabelw{(c) t = 1.030 s}}
        \end{overpic}
        \caption{
            (a) Simulated view of the Phantom v711 fast camera in the configuration used
            for discharge 35628. (b/c) Synchrotron radiation images observed
            using a filtered visible light camera in ASDEX Upgrade during discharge 35628.
            A sudden, sub-millisecond transition from a circular to a crescent
            shape is observed around $t = \SI{1.030}{s}$.
        }
        \label{fig:synchrotronL}
    \end{center}
\end{figure}

For the experiment, a Phantom V711 fast visible camera (connected to the in-vessel optics with an image guide and housed in a shielding box near the tokamak~\citep{yang133d}) was equipped with a narrow-band wavelength filter with central wavelength $\lambda_0 = \SI{708.9}{nm}$ and FWHM of $\SI{8.6}{nm}$. The filter wavelength was chosen as to minimise background line radiation and emphasise the synchrotron radiation, which is emitted in a continuous spectrum and had a higher intensity at longer wavelengths in these plasmas. A simulation of the camera view in discharge \#35628 based on a CAD model is shown in figure~\ref{fig:synchrotronL}a, with details of the configuration presented in table~\ref{tab:camera_parameters}. We note that due to the lack of reliable post-disruption magnetic equilibrium reconstructions, we use the more accurate pre-disruption magnetic equilibrium reconstructions from \CLISTE~\citep{McCarthy1999}  for our synchrotron simulations.

\begin{table}
    \begin{center}
        \begin{tabular}{lc}
            \textbf{Parameter} & \textbf{Value}\\\hline
            Position $(X, Y, Z)$ & $(0.685, -2.300, -0.120)\,\si{m}$\\
            Viewing direction $(n_x, n_y, n_z)$ & $(0.257, 0.964, 0.061)$\\
            Vision angle & $\SI{0.675}{rad}$\\
            Roll angle & $\SI{0.1}{rad}$ (clockwise)\\
            Spectral range & $\SI{708.9}{nm}$ (FWHM $\SI{8.6}{nm}$)\\
            Frame rate & $\SI{1}{kHz}$\\
            Frame resolution & $400\times 400\,$pixels
        \end{tabular}
        \caption{
            Parameters of the image recorded by a Phantom v711 visible light camera, which was
            used for synchrotron radiation imaging. Only parameters relevant
            to synthetic diagnostic simulation are shown.
        }
        \label{tab:camera_parameters}
    \end{center}
\end{table}

\begin{figure}
    \begin{center}
        \includegraphics[width=\textwidth]{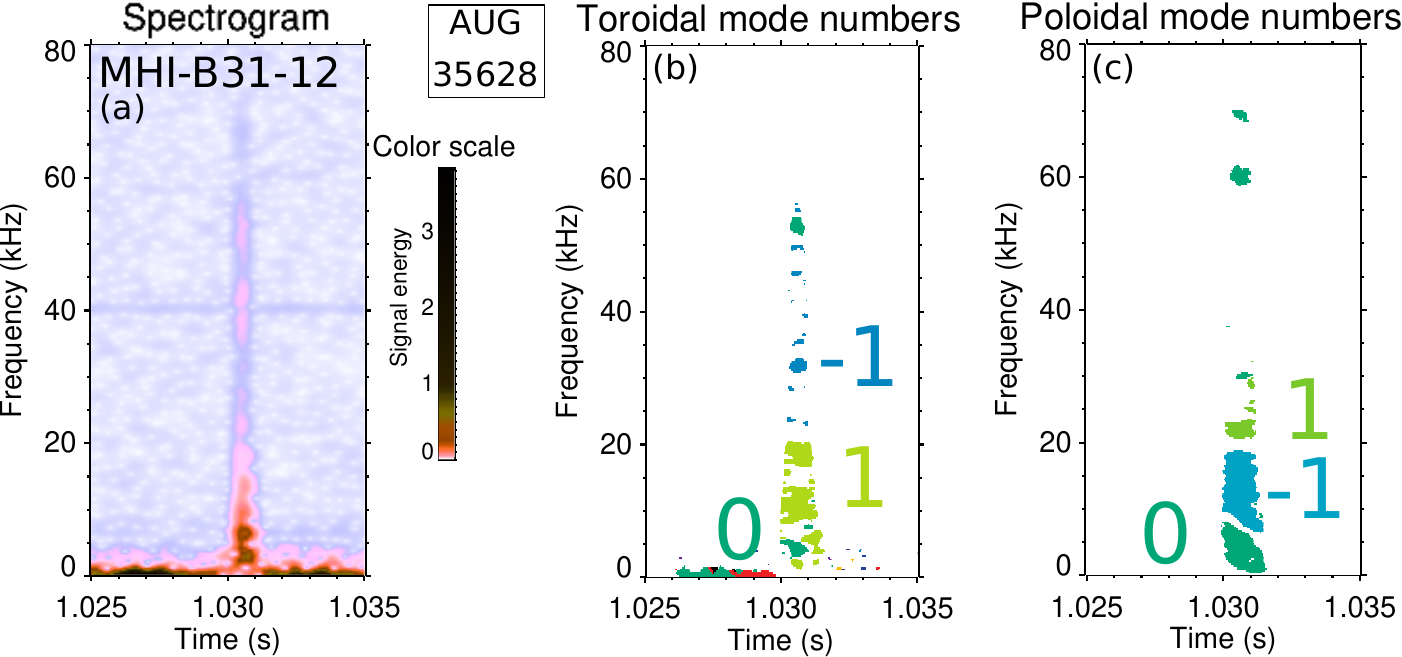}
        \caption{
            Time-frequency analysis of the transient MHD event in the runaway plateau stage of AUG discharge \#35628. (a) Representative spectrogram of a magnetic pick-up coil signal shows wide-band activity with signal energy concentrated to below $\SI{20}{kHz}$; (b) Toroidal mode numbers fitted using the MHI-B31 toroidal ballooning coil array; (c) Poloidal mode numbers fitted using the MHI-C09 poloidal Minrov coil array. Mode number plots show only the good fits and only in regions of sufficient signal energy. The mode below 20 kHz has $(n,m)=(1,-1)$ mode numbers in machine coordinates, which corresponds to $(n,m)=(1,1)$ propagating in the electron diamagnetic drift direction in plasma coordinates.
        }
        \label{fig:mhd-analysis}
    \end{center}
\end{figure}

A few milliseconds after the gas has been injected, a circular synchrotron pattern appears in the visible light camera images. During the next $\sim\SI{20}{ms}$, in the runaway plateau phase of the disruption, the synchrotron pattern is observed to gradually increase in brightness while maintaining approximately the same size and shape. Eventually, the pattern attains its maximum brightness near $t=\SI{1.029}{s}$, shown in figure~\ref{fig:synchrotronL}b. In the very next frame, at $t=\SI{1.030}{s}$ the synchrotron pattern has been turned into a crescent shape, as shown in figure~\ref{fig:synchrotronL}c. Around this time, a modest $\SI{5}{kA}$ spike is observed in the total plasma current, shown in figure~\ref{fig:overview35628}a, corresponding roughly to a $2.5\%$ increase. During this current spike, broadband transient magnetic activity is measured by the magnetic pick-up coils in frequencies ranging from a few kHz to about 80 kHz (see figure~\ref{fig:mhd-analysis}a). Mode number analysis (\ref{fig:mhd-analysis}b-c) was performed using the NTI Wavelet Tools\footnote{https://github.com/fusion-flap/nti-wavelet-tools} program package primarily with a method based on looking for the best fitting integer mode number on the measured cross-phases as function of relative probe positions \citep{horvath}. The toroidal array of ``ballooning coils'', which measures variations to the radial magnetic field, was used for the toroidal mode number analysis, applying the phase corrections of the measured transfer functions of the probes \citep{horvath}. The poloidal mode numbers were determined using the C09 Mirnov coil array, using the probe positions transformed into the straight-field-line coordinate system of the $q=2$ surface. Measured transfer functions were not available for this set of probes, so the confidence in the results had to be improved by applying a complementary method of mode number estimation that is based on the monotonization of the phase functions \citep{Pokol}. The analysis clearly indicated a $(m,n)=(1,1)$ mode propagating in the electron diamagnetic drift direction in the frequency range of 8-$\SI{20}{kHz}$. No precursor activity is observed; the only signal components preceding the event are some low frequency ($\sim\SI{100}{Hz}$) oscillations, which is attributed to vessel and diagnostic vibrations.

%% file: section3.tex
\section{Numerical modelling of the runaway electron distribution}\label{sec:numerics}
A key objective of synchrotron radiation analysis is to validate theoretical models for runaway electron dynamics. Given a runaway electron distribution function from such a model, we should require that the corresponding synthetic synchrotron radiation image matches the experimental image well with regard to pattern shape, size and intensity distribution. Failure to predict the observed synchrotron radiation pattern can provide insight into which effects are missing from the model. In this section, we will discuss the coupling of the 1D fluid code \GO~\citep{Smith2006,Feher2011,Papp2013} to the 2D kinetic solver \CODE~\citep{Landreman2014,Stahl2016}, used to solve for the runaway electron distribution in an ASDEX Upgrade-like disruption.

\subsection{Description of numerical model}
We describe the evolution of the parallel electric field $E_\parallel$ in radius and time by the induction equation in a cylinder, which is solved using \GO:
\begin{equation}\label{eq:go}
    \frac{1}{r} \frac{\partial }{\partial r}\left( r\frac{\partial E_\parallel}{\partial r} \right) = \mu_0\frac{\partial j}{\partial t}.
\end{equation}
Here $r$ denotes the minor radius, $j$ the plasma current density and $\mu_0$ is the permeability of free space. We assume that the plasma is surrounded by a perfectly conducting wall, and that the wall and plasma are separated by a vacuum region that is $\SI{8}{cm}$ wide. The coupling to the kinetic runaway model enters in the current density, $j$, which is decomposed into an Ohmic component $j_\Omega$ and a runaway component $j_{\rm RE}$,
\begin{equation}\label{eq:current}
    j = j_\Omega + j_{\rm RE} = \sigma E_\parallel + e\int v_\parallel f_{\rm RE}\,\dd^3p,
\end{equation}
where $\sigma$ is the electrical conductivity with a neoclassical correction \citep{Smith2006}, $e$ is the elementary charge, $v_\parallel$ is the electron parallel velocity and $f_{\rm RE}$ is the runaway electron distribution function. The distribution function is in turn calculated in every time step, at each radius, using the local plasma parameters, by solving the kinetic equation
\begin{equation}\label{eq:code}
    \frac{\partial f}{\partial t} + eE_\parallel\frac{\partial f}{\partial p_\parallel} =
    C\left\{ f \right\} + S_{\rm ava}.
\end{equation}
The linear collision operator $C\left\{ f \right\}$ accounts for collisions between electrons and (partially screened) ions~\citep{Hesslow2017,Hesslow2018generalized}, and between electrons using a relativistic test-particle operator~\citep{Pike2014}, while the source term $S_{\rm ava}$ accounts for secondary runaway electrons generated through the avalanche mechanism.  We use the simplified avalanche source derived in \citep{Rosenbluth1997} as the difference to the fully conservative operator is small \citep{Embreus2018}.  Radiation losses were initially also considered, but were found to be negligible in this scenario.

The use of a test-particle collision operator in equation~\eqref{eq:code}, given in Appendix A of \citet{Hesslow2019}, is what makes the combined solution of equations~\eqref{eq:go}-\eqref{eq:code} computationally feasible. Neglecting the field particle part of the collision operator, however, also means that the Ohmic current in \CODE\ is underestimated by about a factor of two \citep{HelanderSigmar}, which we must account for in our coupled fluid-kinetic calculations with a self-consistent electric field evolution. The linear relation between $j_\Omega$ and $E_\parallel$ simplifies the correction procedure. By scanning over a wide range of effective charge and temperature, it is found that the conductivity obtained with the test-particle operator is related through a multiplicative factor $g(\Zeff)$ to the fully relativistic conductivity $\sigmaBK$ obtained by~\citet{Braams1989}:
\begin{equation}
    \sigma_{\rm CODE, tp} = g\left(\Zeff\right)\sigmaBK.
\end{equation}
Hence, in order to calculate the runaway contribution to equation~\eqref{eq:current}, we subtract the corrected Ohmic contribution from the total current density $j_{\rm CODE}$ in \CODE:
\begin{equation}
    j_{\rm RE} = j_{\rm CODE} - \sigma_{\rm CODE, tp}E_\parallel = j_{\rm CODE} - g\left(\Zeff\right)\sigmaBK E_\parallel.
\end{equation}
With this approach, the runaway current contribution can be calculated without arbitrarily defining a runaway region in momentum space, while providing a more accurate estimate than assuming all runaways to travel at the speed of light parallel to the magnetic field, which is otherwise usually done in \GO\ and other fluid codes.

To compute the synchrotron radiation observed by the visible camera from the population of electrons calculated using the model above, we use the synthetic diagnostic tool \SOFT~\citep{Hoppe2018a}. This tool calculates, for example, a synchrotron image by summing contributions from all parts of real and momentum space and weighting them with the provided distribution function. To reduce memory consumption, phase space is parameterised using guiding-centre orbits. \SOFT\ can also be used to calculate so-called radiation Green's functions $G(r,p,\thetap)$, as introduced in equation~\eqref{eq:Ihat}, which relate phase-space densities to measured diagnostic signals. This mode of running \SOFT\ is used extensively for the backward modelling in section~\ref{sec:currentprofile}.

\subsection{Plasma parameters used in the numerical simulations}

Runaway electrons in ASDEX Upgrade disruptions, such as \#35628, are typically generated through a combination of the hot-tail and avalanche mechanisms \citep{InsulanderBjork2020}. The avalanche exponentiation factor is robust and depends mainly on the change in the poloidal flux profile. In the ideal theory, where the avalanche growth rate is directly proportional to the electric field, and radial transport during the CQ is assumed to be negligible, the final plateau runaway current profile is completely determined by the surviving post-TQ runaway seed and plasma current profile. In this work we assume that the loop voltage is constant across flux surfaces just before the disruption, and take the initial current profile to be the corresponding ohmic current profile, which leaves the post-TQ seed profile as the main unknown of the simulation. Due to the relatively large current drop of $\sim\SI{600}{kA}$, we expect a significant number of avalanche multiplications to occur throughout the plasma, and therefore a relatively weak sensitivity to the chosen runaway seed density profile.

Simulations with \GO+\CODE\ show that taking into account all the hot-tail electrons obtained from kinetic theory would overestimate the final runaway current with approximately a factor of four. The reason for this is that due to the presence of intense magnetic fluctuations during the TQ, a large part of the hot-tail runaway seed is likely to be deconfined, and the corresponding radial losses are not taken into account in the model. We therefore choose to prescribe a radially uniform seed population such that the final runaway current is matching the experimentally observed plasma current during the plateau. 

With this picture in mind, we take the following approach to modelling discharge \#35628:
\begin{enumerate}[(i)]
    \item First, we perform a purely fluid modelling of the thermal quench using \GO, to obtain the initial electric field evolution. Then, we start the combined kinetic-fluid simulation \emph{after} the thermal quench, thereby effectively disabling the ``natural'' hot-tail generation otherwise obtained in the suddenly cooling plasma. Thus, as an initial condition for the post TQ distribution function, we assume that, at each radius, it consists of current-carrying thermal electrons, along with a smaller electron population, representing the hot-tail that is uniform in radius and Gaussian in the momentum $p$, centred at $p_\parallel=3m_ec, p_\perp = 0$, and with standard deviation $\Delta p = 3m_ec$.
    \item The evolution of the temperature during the TQ itself is taken from the experiment. While the uncertainties of this data are large, we have confirmed that, as a result of prescribing the runaway seed to match the final plasma current, our final results are not sensitive to the details of the temperature evolution. Furthermore, even though the temperature evolution affects the self-consistent electric field evolution during the CQ, the final runaway current is mainly sensitive to the time-integrated electric field, which is independent of temperature evolution. The post-disruption temperature is therefore taken to be $T=\SI{5}{eV}$ throughout the plasma, which is largely consistent with simulated values during the current quench of ASDEX Upgrade disruptions induced by argon gas injection~\citep{InsulanderBjork2020}. Although the temperature is expected to drop to a significantly lower value during the runaway plateau, it only weakly impacts the runaway dynamics and consequently we neglect this effect here. 
    \item We assume that neutral argon atoms with density $n_{\rm Ar} = \SI{0.83e20}{\per\metre\cubed}$ (corresponding to 20\% of the injected atoms \citep{Pautasso2020,InsulanderBjork2020}) are uniformly distributed in radius at the beginning of the simulation, and neglect their radial transport throughout the simulation. The densities of the various ionization states, and the corresponding electron density, are calculated assuming an equilibrium between ionization and recombination; that is, $n^{i}_k$ are computed from
\begin{equation}
   \begin{aligned}
    &R^{i+1}_k n^{i+1}_k-I^i_k n^i_k=0,  \hspace{0.5cm}  i=0,1,..., Z-1,\\
    &\sum_i n^{i}_k=n^\mathrm{tot}_k,  \hspace{0.5cm} i=0,1,..., Z,
    \end{aligned}
    \label{eq:ioniz_equilibrium}
\end{equation}
where $n^\mathrm{tot}_k$ is the total density of species $k$.
$I_k^{i}$ denotes the electron impact ionization rate and $R_k^i$ 
the radiative recombination rate for the $i^{\rm th}$ charge state of
species $k$, respectively. The ionization and recombination rates are extracted from the Atomic Data and Analysis Structure (ADAS) database~\citep{Summers2007,Summers2011}.

\end{enumerate}

\subsection{Simulation results}\label{sec:numerics:results}

\begin{figure}
    \centering
    \includegraphics[width=0.8\textwidth]{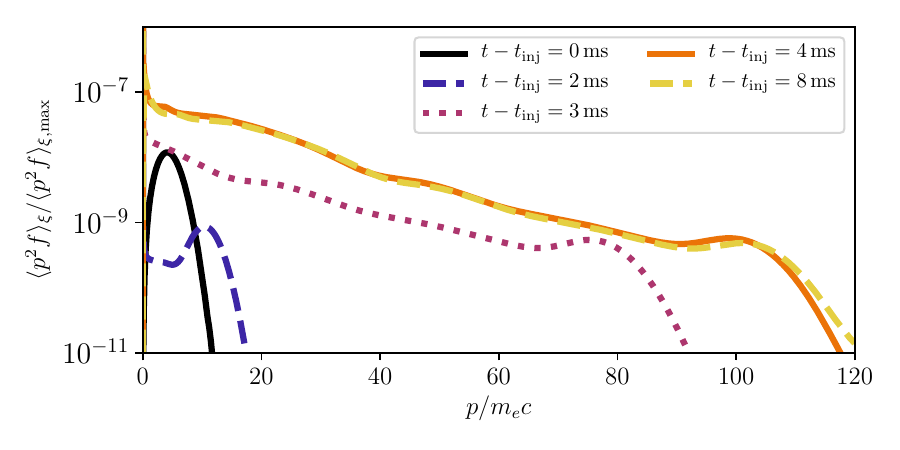}
    \caption{
        Time evolution of the electron energy spectrum (pitch-averaged distribution function) at the magnetic axis. The runaway electron seed starts close to $p=5m_ec$ at $t=0$, and is then quickly accelerated to above $p=100m_ec$ within a few milliseconds. During the remainder of the runaway plateau, the initial seed sits around $p=100m_ec$ while new runaway production is dominated by large-angle collisions, causing the energy spectrum to slowly approach an exponential. The distribution also contains a thermal Maxwellian component, but due to its low temperature, it only appears as a vertical line at $p=0m_ec$ in this figure.
    }
    \label{fig:REenergy}
\end{figure}

The time evolution of the electron energy spectrum in the \GO+\CODE\ simulation of discharge \#35628 is shown in figure~\ref{fig:REenergy}. We only evolve the simulation until around $t=\SI{1.029}{s}$, before the synchrotron pattern suddenly changes. The only plausible mechanism that could cause such a rapid transition of the pattern is a relaxation of the current profile in fast magnetic reconnection~\citep{Igochine2006,Papp2011}, in relation to an internal magnetohydrodynamic instability---a physical mechanism beyond the modelling capabilities of the \GO+\CODE\ tool.

As shown in figure~\ref{fig:REenergy}, the seed runaway population is quickly accelerated to a maximum energy during the current quench, which lasts for approximately $\SI{4}{ms}$. During this phase, a population of secondary runaways gradually builds up, overtaking the plasma current. The maximum energy varies across radii---from $p\approx 100m_ec$ in the core to only a few $m_ec$ at the edge---as it primarily depends on the magnitude of the induced electric field during the disruption, which, in turn, depends on the change in the current profile; in this scenario the maximum energy decreases monotonically with radius from its maximum value on the magnetic axis. After the current quench, during the remainder of the simulation, the seed electrons remain around this maximum energy as the electric field has dropped to the low level required to sustain the runaway current.

\begin{figure}
    \centering
    \begin{overpic}[width=0.24\textwidth,trim={55mm 45mm 22mm 30mm},clip]{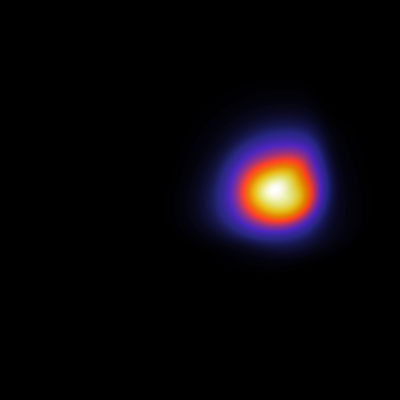}
        \put(5,90){\figlabelw{(a) \SOFT}}
        \put(5,81){\figlabelw{$t = \SI{1.008}{s}$}}
    \end{overpic}
    \begin{overpic}[width=0.24\textwidth,trim={55mm 45mm 22mm 30mm},clip]{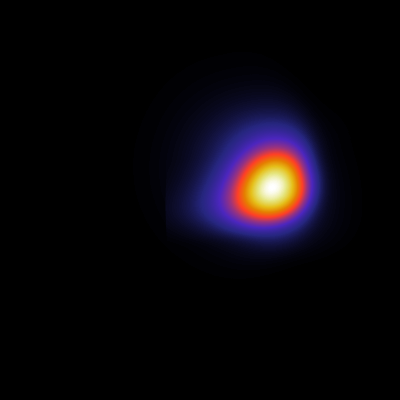}
        \put(5,90){\figlabelw{(b) \SOFT}}
        \put(5,81){\figlabelw{$t = \SI{1.018}{s}$}}
    \end{overpic}
    \begin{overpic}[width=0.24\textwidth,trim={55mm 45mm 22mm 30mm},clip]{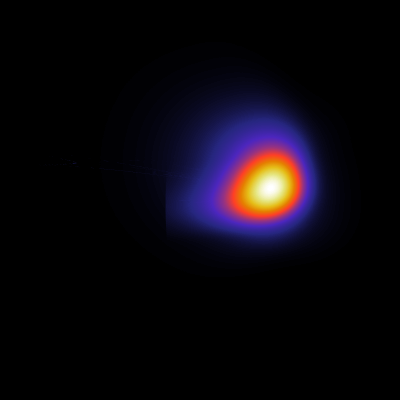}
        \put(5,90){\figlabelw{(c) \SOFT}}
        \put(5,81){\figlabelw{$t = \SI{1.029}{s}$}}
    \end{overpic}
    \begin{overpic}[width=0.24\textwidth,trim={55mm 45mm 22mm 30mm},clip]{figure3b.png}
        \put(5,90){\figlabelw{(d) Experiment}}
        \put(5,81){\figlabelw{$t = \SI{1.029}{s}$}}
    \end{overpic}
    \caption{
        Comparison of the synthetic synchrotron image produced with \SOFT, taking the distribution function calculated with \GO+\CODE\ at (a) $t=\SI{1.008}{s}$, (b) $\SI{1.018}{s}$, and (c) $\SI{1.029}{s}$ as input, with the (d) synchrotron image taken in ASDEX-U \#35628, also at $t=\SI{1.029}{s}$. Although the synthetic synchrotron pattern is larger than the experimental pattern, the overall shape of the two patterns is the same, indicating that the overall runaway dynamics are well explained by \GO+\CODE.
    }
    \label{fig:gocodeimage}
\end{figure}
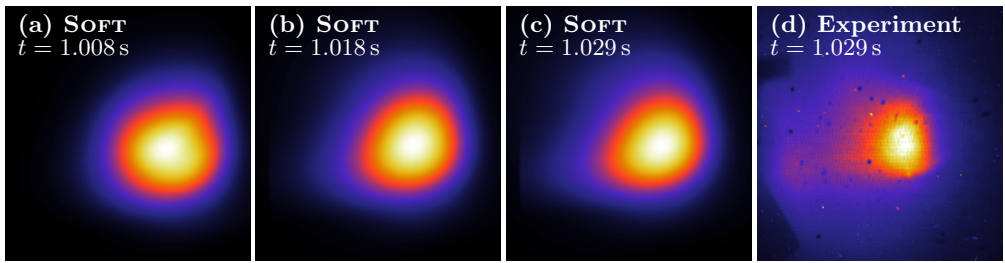

Using \SOFT, we compute the synchrotron radiation observed from the distribution of electrons calculated with \GO+\CODE. The resulting synthetic camera image at $t=\SI{1.029}{s}$, just before the pattern transition, is presented in figure~\ref{fig:gocodeimage}c (along with two preceding times in figure~\ref{fig:gocodeimage}a-b). Comparing the synthetic image to the experimental image in figure~\ref{fig:gocodeimage}d, we find qualitative agreement, with both synchrotron patterns taking a round shape. The synthetic pattern is, however, significantly larger than the experimental pattern, both horizontally and vertically. The size of the synchrotron pattern is directly related to the radial density of runaway electrons, suggesting that the radial runaway density profile is more sharply peaked in the experiment than in the \GO+\CODE\ simulation, which has a flat runaway density profile. An explanation for this discrepancy could be that the assumed runaway seed profile differs from that in the experiment, or that the initial current profile---which would experience flattening during the current spike in the thermal quench---is different. Deviations in plasma parameters such as the density and temperature could also have an effect on the avalanche gain during the current quench, as could the radial transport, which we do not model. All of these are assumed parameters in our model, due to the lack of low uncertainty experimental data. These parameters could be adjusted to give a better matching radial distribution of synchrotron radiation. 
However, improving agreement this way would be both computationally expensive and of limited value in better illuminating the underlying physics, so we leave this exercise for future studies.  

\begin{figure}
    \centering
    \hfill\includegraphics[width=0.95\textwidth]{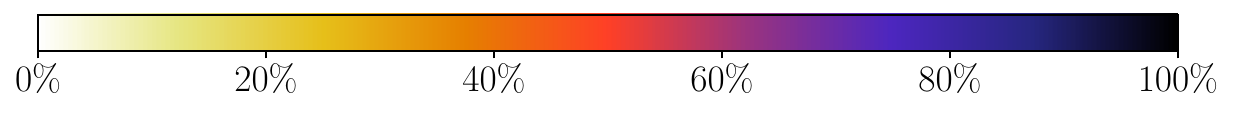}
    \includegraphics[width=\textwidth]{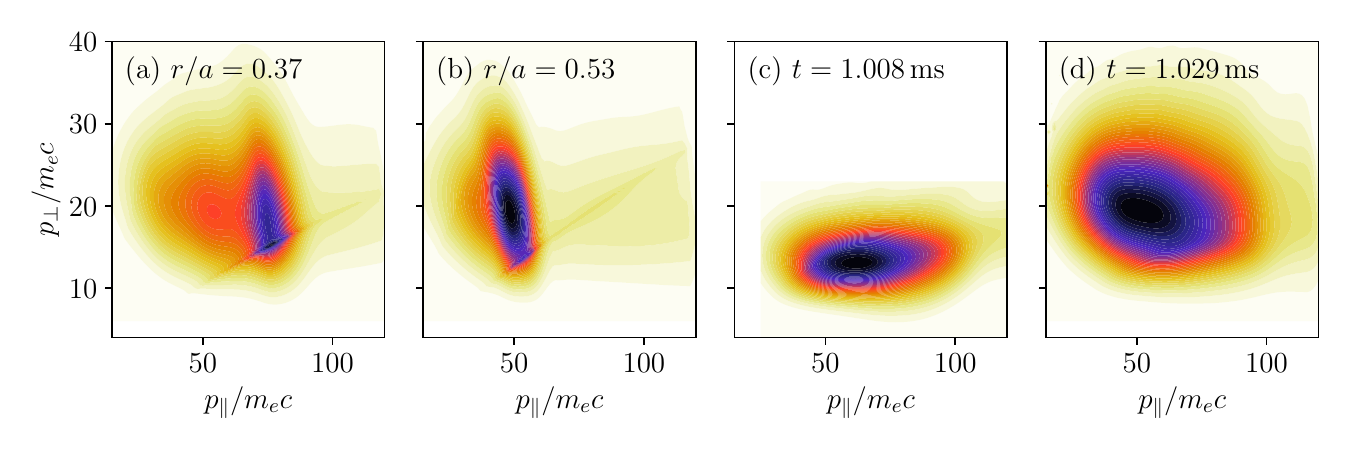}
    \caption{Amount of synchrotron radiation observed from different parts of momentum space. Panels (a) and (b) show the contributions at $t=\SI{1.029}{ms}$ from two individual radii, indicating that the emission is dominated by the remnant hot-tail seed. The very sharp features running along almost constant pitch in (a) and (b) are physical, and are connected to the very bright edges usually seen in synchrotron images from mono-energetic and mono-pitch distribution functions. Panels (c) and (d) compare the radially integrated synchrotron radiation at $t=\SI{1.008}{ms}$ and $t=\SI{1.029}{ms}$ respectively.}
    \label{fig:gocodesoft_super}
\end{figure}

\begin{figure}
    \centering
    \includegraphics[width=\textwidth]{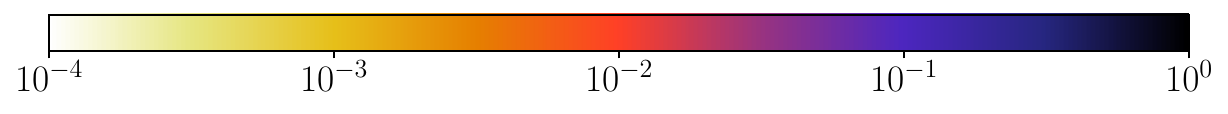}
    \includegraphics[width=\textwidth]{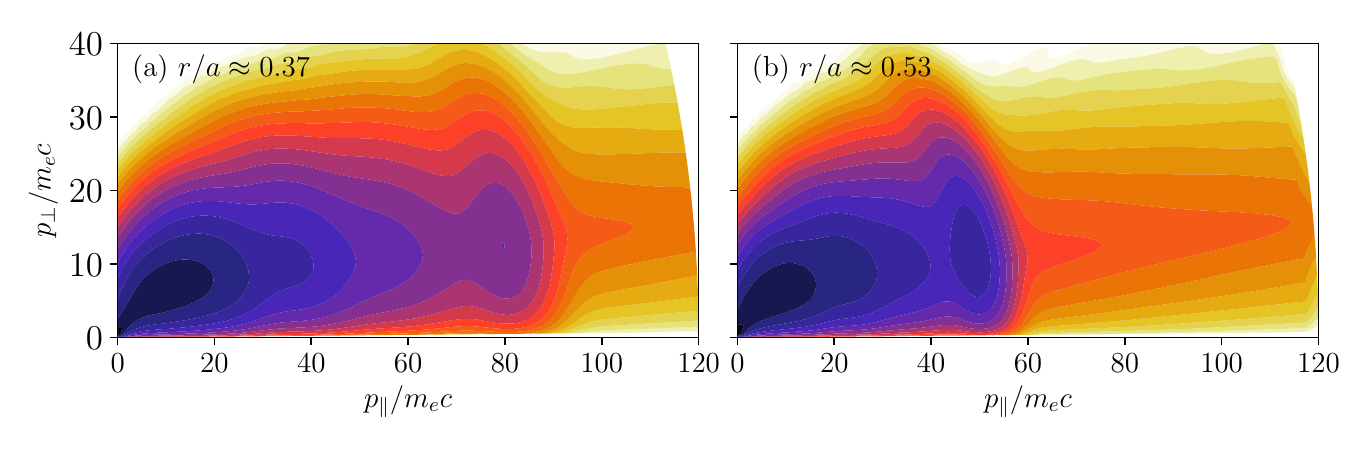}
    \caption{Momentum space distribution functions (multiplied by the momentum-space Jacobian $p^2\sin\thetap$) from the \GO+\CODE\ simulations at two select radii, chosen to correspond approximately to the particles contributing to figures~\ref{fig:gocodesoft_super}a and~\ref{fig:gocodesoft_super}b. The remnant seed appears as a bump in the distribution function around (a) $p_\parallel = 80 m_ec$ and (b) $p_\parallel = 55m_ec$.}
    \label{fig:gocodesoft_dist2d}
\end{figure}

Instead, we turn our attention to the source of the observed radiation; a closer analysis reveals that most of it is emitted by an ensemble of particles that originally constituted the hot-tail seed. As was shown in figure~\ref{fig:REenergy}, these electrons were rapidly accelerated during the current quench and then remained at their peak energy. In figures~ \ref{fig:gocodesoft_super}a and b, we show the synchrotron radiation observed from the particles associated with the $r/a = 0.37$ and $r/a = 0.53$ flux surfaces, respectively. By comparing the origin of the radiation at these radii with the local momentum space distributions in figures~\ref{fig:gocodesoft_dist2d}a and b respectively, we find that the region of momentum space that dominates synchrotron emission at each radius coincides with the location of the local seed population. Hence we conclude that it is the remnant seed runaways that dominate synchrotron radiation in these simulations.

\begin{figure}

    \centering
    \includegraphics[width=\textwidth]{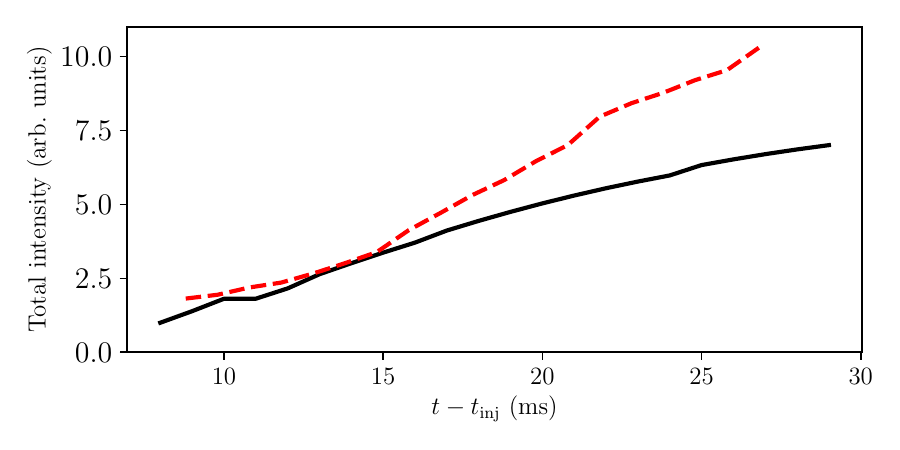}
    \caption{
        Total detected synchrotron intensity as predicted by combined \GO+\CODE\ and \SOFT\ simulations (black, solid), and as recorded in the experiment (red, dashed). Since the experimental measurements are not absolutely calibrated, the curves have been re-scaled to aid comparison of the slopes.
    }
    \label{fig:synchIntensity}
\end{figure}

When integrating over all radii, a wider dominant region appears in momentum space, as shown in figures~\ref{fig:gocodesoft_super}c and d for an early and a late simulation time, respectively. A comparison of the emission at the two times reveals that the dominant region moves towards greater perpendicular momentum as time passes. This is caused by collisional pitch-angle scattering which increases the average perpendicular momentum in the distribution. As a result of the increased perpendicular momentum, the runaways emit more synchrotron radiation, leading to a gradual increase of the total intensity in the camera images. The change in pitch-angle, however, is sufficiently small to not affect the synchrotron pattern shape significantly. Figure~\ref{fig:synchIntensity} shows the time evolution of the total intensity in the simulated (solid black line) and the experimental (dashed red) images, respectively. Although both intensities increase steadily, they do so at slightly different rates. This can be explained by a discrepancy in the argon density used for the simulations. As we show in appendix~\ref{app:kinetic theta}, kinetic theory predicts that electrons with momentum $p$ have an exponential pitch-angle dependence in the disruption plateau phase, $f_\xi(\xi)\sim\exp(C\xi)$, with $C$ a time-dependent constant, and $\xi=p_\parallel/p$. During the runaway plateau, $C$ is roughly inversely proportional to time until it reaches an equilibrium value of $0.1p$. In appendix~\ref{app:kinetic theta} we show that the pitch parameter $C$ evolves approximately as
\begin{equation}
    C(t)\approx\frac{(p/m_ec)^2}{8n_{\rm Ar,20}(t-t_0)_{\rm ms}},
\end{equation}
where $n_{\rm Ar,20}$ is the argon density, measured in units of $\SI{e20}{\per\metre\cubed}$, and the times in the denominator are given in milliseconds. The emitted synchrotron power at a frequency $\omega$ can furthermore be approximated by the contribution from the strongest emitting particle of such a distribution which is proportional to
\begin{equation}\label{eq:Poft}
    \mathcal{P} = \exp\left[ -\left( \frac{\omega m_e}{3\sqrt{2}eB}\right)^{2/3}\frac{1}{\gamma^{\star 4/3}}\frac{1}{\left[\frac{1}{C(t_0)} + (t-t_0)\nu_D \right]^{1/3}}\right],
\end{equation}
with $\gamma^\star = \sqrt{1+(p^\star/m_ec)^{2}}$, and $\pstar$ the momentum reached by the hot tail seed. The free parameters in this expression are $\gamma^\star$, $n_{\rm Ar}$, and the unknown prefactor, and it should therefore in principle be possible to fit this expression to the curves in figure~\ref{fig:synchIntensity}, assuming a constant background plasma parameters and no radial transport. Unfortunately, however, such a fit can be rather ill-conditioned when the data is nearly linear, as is the case here. This is partly due to the relatively short time before the transition in the synchrotron patterns happens, which does not allow for significant pitch angle relaxation. Therefore, in practice, it is not possible to extract a reliable estimate of $n_{\rm Ar}$  in this case. Nevertheless, the ability for~\eqref{eq:Poft} to fit both curves in figure~\ref{fig:synchIntensity} lends credibility to the physical picture obtained from \GO+\CODE\ and may be used to estimate the impurity density in other experiments in the future.

%% file: section4.tex
\section{Backward modelling}\label{sec:currentprofile}
The fluid-kinetic model described in section~\ref{sec:numerics} appears to capture the runaway evolution during the first part of the runaway plateau phase fairly well, but it does not contain the physics necessary to describe the sudden synchrotron pattern transition occurring in the experiment at $t\approx\SI{1.030}{s}$. In this section, we instead analyse the synchrotron radiation images directly and extract information from the camera images using a regularised, direct inversion. Without further constraints a direct inversion of the synchrotron image would be an ill-posed problem, thus we derive an analytical model for the dominant part of the runaway electron distribution building on the results of section~\ref{sec:numerics:results}, which allows us to better constrain runaway parameters.

\subsection{Inversion procedure}
We may capitalise on what we have learned from the fluid-kinetic simulations of section~\ref{sec:numerics}, in order to find constraints to regularise the inversion of the synchrotron images.  
We found that the synchrotron radiation pattern seen in the camera images was dominated by the remnant runaway electron seed population, situated at some maximum energy and slowly relaxing  towards a steady-state pitch distribution. This evolution suggests an accelerated seed electron distribution function of the form
\begin{equation}\label{eq:seedf}
    f_{\rm seed}(r,p,\xi) = f_r(r)\exp\left[ -\left( \frac{p-\pstar}{\Delta p} \right)^2 \right]\exp(C\xi)\approx
    f_r(r)\delta\left( p-\pstar \right)\exp\left( C\xi \right),
\end{equation}
where $f_r(r)$ is an arbitrary function describing the radial runaway density, and $\pstar$, $\Delta p$ and $C$ are free fitting parameters. Approximating the Gaussian with a delta function is motivated by the fact that the synchrotron pattern is usually rather insensitive to variations in the runaway energy distribution. The same argument is used to fully decouple the spatial coordinate $r$ from the momentum parameter $\pstar$ and, since $C$ is mainly a function of $\pstar$, also from $C$. In this model we thus assume for simplicity that the momentum and pitch distributions are the same across all flux surfaces.

Our inversion method utilises the capability of \SOFT\ to generate weight functions for a given tokamak/detector setup, as described in section~\ref{sec:images}. The brightness of pixel $i$ in a synchrotron image $I_i$ is related to the distribution function $f(r,p,\xi) = f_r(r)f_{p\xi}(p,\xi)$ through the weight function $G(r,p,\xi)$ as
\begin{equation}\label{eq:imageeq}
    I_{i} = \int G_{i}(r,p,\xi) f(r,p,\xi)\,p^2\,\dd r\,\dd p\,\dd\xi.
\end{equation}
By representing the image and the discretised radial runaway density profile as vectors, the discretised version of the equation system~\eqref{eq:imageeq} can be formulated as
\begin{equation}
    I_{i} = \sum_k \tilde{G}_{ik}(r) f_r^k(r),
\end{equation}
where we have introduced the reduced weight function matrix
\begin{equation}
    \tilde{G}_{ik} = \Delta r_k \int G_{i}(r_k,p,\xi)f_{p\xi}(p,\xi)\,p^2\,\dd p\dd\xi,
\end{equation}
with $\Delta r_k=|r_{k+1}-r_{k}|$. Given a momentum-space distribution function $f_{p\xi}(p,\xi)$, which can be characterised with the parameters $\pstar$ and $C$ in equation~\eqref{eq:seedf}, we thus seek to minimise the sum of squares of differences between pixels in the synthetic and experimental images $I_i$ and $I_i^{\rm exp}$. Since the problem is still ill-posed, we regularise it using the Tikhonov method~\citep{NumericalRecipes}, resulting in
\begin{equation}\label{eq:TikhonovProblem}
    \begin{aligned}
        f_r(r) 
        &= \arg\min_{f_r}\left[ \left\lVert \sum_j\alpha\Gamma_{ij}f_r^j \right\rVert_2^2 + \sum_i\left\lVert I_i^{\rm exp} - I_i \right\rVert_2^2 \right] =\\
        &=
        \arg\min_{f_r} \left[ \left\lVert \sum_j \alpha\Gamma_{ij}f_r^j \right\rVert_2^2 + \sum_i\left\lVert I_i^{\rm exp} - \sum_j \tilde{G}_{ij} f_r^j \right\rVert_2^2 \right],
    \end{aligned}
\end{equation}
where the Tikhonov matrix $\Gamma_{ij}$ is taken to be the identity matrix and the Tikhonov parameter $\alpha$ is determined using the L-curve method~\citep{Hansen1993}. The ability to solve this problem using a linear least-squares method allows us to efficiently explore the space of possible combinations of $(\pstar, C)$, and hence to use typical minimization methods to solve the full problem.

\subsection{Inverted distribution function}
One of the main reasons for applying backward modelling to this ASDEX-U discharge is to better understand what gives rise to the synchrotron pattern transition between $t=\SI{1.029}{ms}$ and $t=\SI{1.030}{ms}$, corresponding to the video frames~\ref{fig:synchrotronL}b and c. To see that the transition is \emph{not} due to a change of the energy distribution, we can estimate the energy gained by the runaways in a magnetic reconnection event with the following argument: if the plasma current changes by an amount $\Delta I$ during the event, then it follows from a circuit equation for the plasma that the energy of the electrons changes according to
\begin{equation}
    \Delta W = \int ecE\,\dd t =
    -\int \frac{ecL}{2\pi R_0}\frac{\dd I}{\dd t}\,\dd t =
    -\frac{ecL}{2\pi R_0}\Delta I,
\end{equation}
where $L$ is the self-inductance and $R_0$ is the major radius of the torus. Assuming $L\sim \mu_0 R_0$, with $\mu_0$ the vacuum permeability, the induced energy is found to increase by $\SI{60}{eV}$ for every ampere decrease in the plasma current. In our case, where the second current spike rises by $\Delta I\approx\SI{5}{kA}$, the energy transferred to the electrons should be below $\SI{0.5}{MeV}$, which is small compared to typical runaway energies at $t=\SI{1.029}{ms}$ ($\approx\SI{25}{MeV}$ at mid-radius, see figure~\ref{fig:gocodesoft_dist2d}b). Furthermore, since the energy is gained exclusively in the parallel direction, and the pitch angles are initially small, the pitch angles should also be negligibly affected; 
$\Delta\theta\approx -(\Delta W_\| / W_\|) \theta$.

A more general argument against a change of the momentum-space distribution is that the synchrotron intensity is more sensitive to changes in the energy and pitch angle than to changes in the radial density profile. As discussed in section~\ref{sec:images}, the observed synchrotron intensity is exponentially sensitive to $p$ and $\xi$ in the short wavelength limit, whereas the radial density always appears as a multiplicative factor. Since the synchrotron intensity does not change significantly in the spot shape transition, the change to the momentum-space distribution should not be significant either. On the other hand, a parameter scan indicates that a significant change to the momentum-space distribution would be required for a visible spot shape transition.
Hence, we expect the observed synchrotron pattern transition to be caused by a spatial redistribution of runaways. In what follows, we will therefore seek the best fit between theory and experiment for both frames~\ref{fig:synchrotronL}a and b simultaneously, assuming $\pstar$ and $C$ to remain unchanged in the transition.

\begin{figure}
    \centering
    \includegraphics[width=\textwidth]{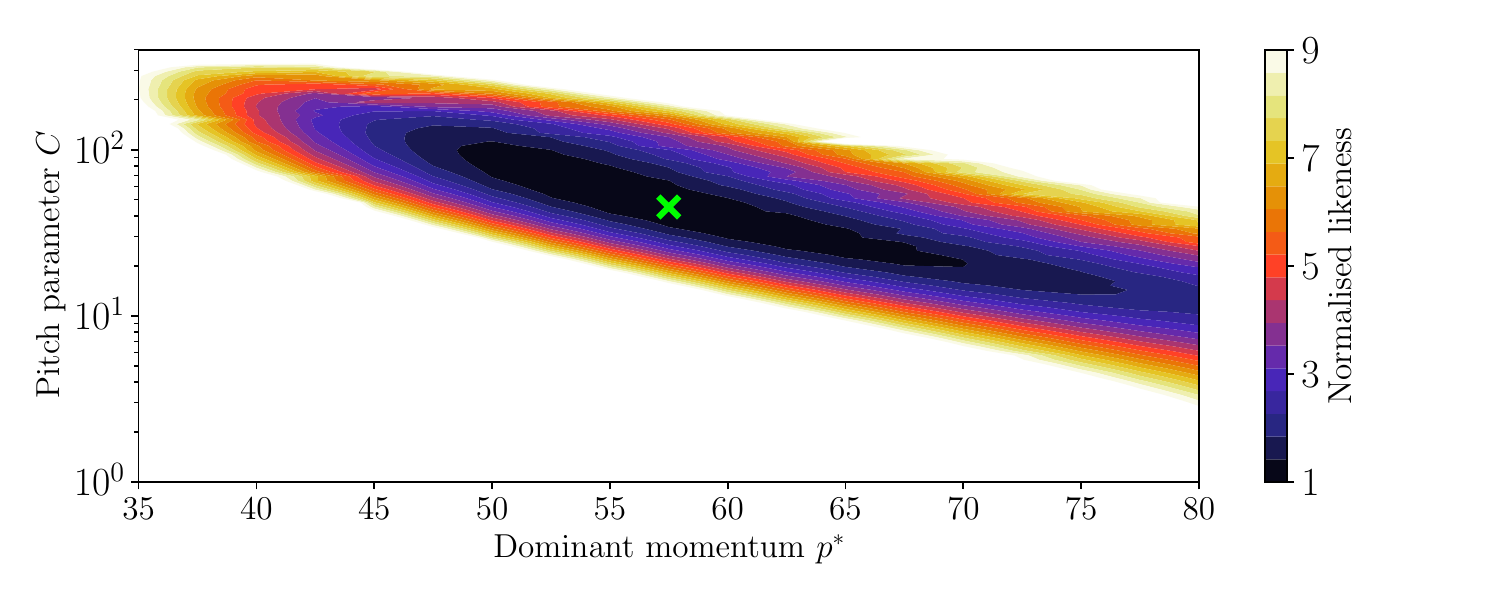}
    \caption{
        Sum of pixel differences squared for both of frames~\ref{fig:backward:frames}a and b, given different combinations of $\pstar$ and $C$. For each combination of $\pstar$ and $C$, the corresponding optimal radial density is calculated and used for comparing the images. The global minimum is marked with a green cross and is located in $\pstar = 57.5$ and $C = 45$. The white regions correspond to unreasonable combinations of the two parameters.
    }
    \label{fig:fitness}
\end{figure}

The sum of pixel differences squared, as a function of the fitting parameters $\pstar$ and $C$, is shown in figure~\ref{fig:fitness}. In each point, the best radial density is constrained using equation~\eqref{eq:TikhonovProblem}. Optimal agreement is obtained with $\pstar=57.5m_ec$ and $C=45$, although the region of good agreement in figure~\ref{fig:fitness} is fairly large. However, most of the optimal combinations of $p$ and $C$ yield approximately the same value for the dominant pitch angle, $\tstar\approx\SI{0.30}{rad}$. This is in agreement with the \GO+\CODE\ and \SOFT\ simulations presented in section~\ref{sec:numerics:results} which had $\pstar\approx 55m_ec$, $C\approx 25$ (at $p=\pstar$) and $\tstar\approx \SI{0.36}{rad}$ at $t=\SI{1.029}{ms}$. Since $C$ is inversely proportional to the (relatively poorly diagnosed) argon density $n_{\rm Ar}$, these are well within the uncertainties of the inversion and in the plasma parameters.

\begin{figure}
    \centering
    \begin{minipage}{0.49\textwidth}
        \begin{overpic}[width=\textwidth]{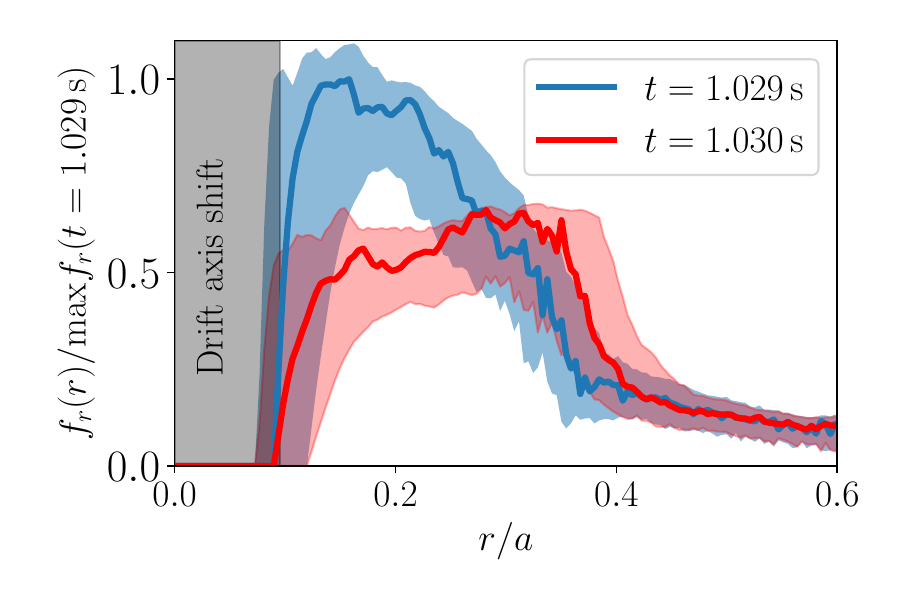}
            \put(22,57){\figlabel{(a)}}
        \end{overpic}%
    \end{minipage}
    \begin{minipage}{0.49\textwidth}
        \begin{overpic}[width=0.49\textwidth]{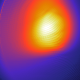}
            \put(5,7){\figlabelw{(b) $t = \SI{1.029}{s}$}}
        \end{overpic}
        \begin{overpic}[width=0.49\textwidth]{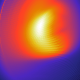}
            \put(5,7){\figlabelw{(c) $t = \SI{1.030}{s}$}}
        \end{overpic}\\
        \vspace{4mm}
    \end{minipage}
    \caption{(a) Inverted radial density profiles for the video frames at $t=\SI{1.029}{s}$ (black) and $t=\SI{1.030}{s}$ (red) for the best fitting values of ($\pstar,C)$, and the corresponding inverted synthetic synchrotron radiation images at (b) $t=\SI{1.029}{s}$ and (c) $\SI{1.030}{s}$. The optimal values of $\pstar$ and $C$ extracted from figure~\ref{fig:fitness} and used to generate the images are $\pstar = 57.5m_ec$ and $C\approx 45$. The blue and red shaded regions in (a) indicate the maximum variation of the radial profiles among all solutions with normalised likeness $\leq 2$ (corresponding to all points within the 2 contour of figure~\ref{fig:fitness}). The grey shaded region has the size of the drift orbit shift and contains no particles since \SOFT\ only counts particles in the \emph{outer} midplane.
    }
    \label{fig:backward:frames}
\end{figure}

The radial density profiles obtained in the inversion for the two video frames are presented in figure~\ref{fig:backward:frames}a. Note that the radial coordinate denotes the particle position along the outer midplane, and that $r=0$ corresponds to the magnetic axis. Since particles are counted on the outboard side, and since only particles with $p=\pstar$ are considered, the inverted radial profile is exactly zero within the grey region $r/a\approx 0.1$, corresponding to the drift orbit shift for these particles. The blue and red shaded regions in figure~\ref{fig:backward:frames}a indicate the maximum deviation of density profiles corresponding to the optima of all combinations $(\pstar, C)$ with normalised likeness less than 2 in figure~\ref{fig:fitness}. Since the radial density profiles contain an uninteresting scaling factor in order for the inversion algorithm to match the absolute pixel values in both experimental and synthetic images, we rescale all density profiles by a scalar multiplicative factor before evaluating the maximum deviation. The maximum deviations suggest that although the uncertainty in $(\pstar,C)$ is relatively large, the radial density profiles are somewhat more robust. Specifically, the analysis shows that the synchrotron pattern transition must be due to a spatial redistribution of particles.

The radial density profile inversion reveals that the synchrotron pattern change is caused by a rapid redistribution of the runaway electron density. As described in section~\ref{sec:images}, the shifted density profile causes the particles on the high-field side to collectively emit relatively more radiation than those on the low-field side, yielding a crescent pattern after the event. By integrating the radial density profiles in figure~\ref{fig:backward:frames}a over the tokamak volume, one finds that approximately $14\%$ of particles are predicted to be lost in the event.

A redistribution of the runaway density profile is also consistent with the occurrence of magnetic reconnection, which, in addition to causing the observed current spike, would redistribute the current profile. We do not have sufficient experimental information to perform an MHD stability analysis, thus it is uncertain exactly how this reconnection event is triggered. However, the fact that runaway generation is often more efficient in the plasma core \citep{Eriksson2004}, it is plausible that a corresponding increase in the peaking of the current profile and decreasing central safety factor, could destabilize an internal kink instability, consistent with the strong (1,1) mode activity evident in \ref{fig:mhd-analysis}b. 

%% file: section5.tex
\section{Summary}\label{sec:conclusions}
We analyzed visible-light camera images of synchrotron emission from runaway electrons in the ASDEX Upgrade discharge 35628 that was deliberately disrupted by argon massive gas injection.  In the runaway plateau phase of this particular discharge, a sudden transition of the synchrotron image from circular to crescent shape was observed, correlated with a spike in the plasma current as well as $(m,n)=(1,1)$ MHD activity.  

We simulated the spatio-temporal evolution of the runaway distribution function with a coupled kinetic-fluid code taking into account the evolution of the electric field self-consistently. This distribution function was then used as an input to the \SOFT\ synthetic diagnostic tool to calculate the shape and intensity evolution of the synchrotron images. We find that a hot-tail seed population of runaway electrons was multiplied by close collisions, resulting in a distribution with an exponentially decreasing energy spectrum.   The remnant seed was accelerated to high energies and was responsible for most of the synchrotron emission. 

We derived an analytical expression for the time evolution of the total emitted synchrotron power and found qualitative agreement with the observed intensity evolution. By constraining the distribution function in momentum space we inverted the radial profiles of the runaway beam and found that the change from circular to crescent shape was caused by a redistribution of the radial profile of the RE density. In particular the momentum-space distribution was found to be difficult to extract from the images, and to better constrain it one could potentially combine several independent diagnostic signals, such as synchrotron spectra, bremsstrahlung spectra, and camera images that view the plasma from different angles, or are sensitive to different wavelength ranges.
 
A number of directions for future studies could be envisaged based on the results of this study. Radial transport is expected to significantly affect runaway dynamics, both in present-day as well as future tokamaks, however this has often been neglected in runaway models due to its complexity, including in this work. Finite-aspect-ratio effects have also been neglected in the fluid-kinetic modelling part of this study, but could play a role in the evolution of the runaway electron distribution function. Finally, the ``two-component'' runaway population observed in ASDEX Upgrade discharge \#35628, studied here, consisting of a high-energy remnant seed and a current-carrying avalanche component, should be investigated in different devices and discharges. If the situation observed in ASDEX Upgrade is representative for many devices, the two-component model used here for fitting synchrotron radiation could be used more widely to infer runaway parameters at many experiments.

%% file: appendixA.tex
\section{Accounting for guiding-centre drifts}\label{app:drifts}
To reduce the size of the phase-space from six to three dimensions, \SOFT\ utilises a guiding-centre transformation which eliminates the orbit time, as well as the toroidal and gyro-phase angles from the distribution function. However, the dependence on the particle's position and momentum components still remains in the radiation emission function, $\dd P/\dd \Omega$; these must be computed within the guiding-centre framework. The runaway electron drift orbit shift can however be significant, and so the effect is of importance in many, or even most, practical scenarios. In this section we describe the incorporation of guiding-centre orbit drifts in \SOFT.

\subsection{Electron momentum vector}
Formally, the guiding-centre drift motion appears in the standard theory as an $\mathcal{O}(\epsilon)$ effect, where $\epsilon$ is the usual guiding-centre ordering parameter. An expression for the Larmor radius vector to first order was given by~\cite{Hirvijoki2015}, and using expressions found therein one can also derive the following expression for the momentum vector of a particle with mass $m$ and charge $q$ (negative for electrons):
\begin{equation}\label{eq:momvec}
    \begin{aligned}
        \bb{p} &= \bb{P} +
        \pperp\mathrm{sign}(q)\perphat
        + \epsilon\bhat\left[ \ppar\rho\left(\rhohat\cdot\bb{\kappa} \right) + \frac{m\mu}{q}\left(\aur_1 : \nabla\bhat \right) \right] +\\
        &+ \epsilon\rhohat\left[  qB\frac{\rho^2}{2}\left( \perphat\cdot\nabla\ln B\right) - \frac{\rho\ppar}{2}\left( \bhat\cdot\nabla\ln B\right) - \rho\ppar\left(\aur_2 : \nabla\bhat \right)\right] +\\
        &+ \epsilon\perphat\left[ qB\frac{\rho^2}{2}\left(\rhohat\cdot\nabla\ln B\right) - \frac{\rho\ppar}{2}\left(\aur_1 : \nabla\bhat\right) \right] + \mathcal{O}(\epsilon^2),
    \end{aligned}
\end{equation}
with the guiding-center momentum vector
\begin{equation}
    \bb{P} = \ppar\bhat + \epsilon\frac{\bhat}{qB^\star_\parallel}\times
    \left(m\mu \nabla B + \ppar^2\bb{\kappa} \right),
\end{equation}
where $\ppar$ and $\pperp$ are the particle momenta parallel and perpendicular to the magnetic field respectively, $B$ is the magnetic field strength, $\bhat$ and $\rhohat$ are mutually perpendicular unit vectors in the direction of the magnetic field and (lowest order) Larmor radius vector, respectively, $\perphat = \rhohat\times\bhat$, $\rho = \pperp/|q|B$ is the Larmor radius, $\mu=\pperp^2/2mB$ is the magnetic moment, $\bb{\kappa} = (\bhat\cdot\nabla)\bhat$ is the inverse curvature vector, $B^\star_\parallel = B+\ppar\bhat\cdot(\nabla\times\bhat)/q$, and the dyads $\aur_1$ and $\aur_2$ are defined as
\begin{equation}
    \aur_1 = -\frac{1}{2}\left( \rhohat\perphat + \perphat\rhohat \right), \qquad
    \aur_2 = -\frac{1}{4}\left( \perphat\perphat - \rhohat\rhohat \right).
\end{equation}
To lowest order, the momentum vector~\eqref{eq:momvec} describes circular gyro motion around the magnetic field line. However, when including $\mathcal{O}(\epsilon)$ terms, the gyro motion component picks up contributions which alter this circular motion. A detailed analysis reveals that the first-order terms primarily shift the axis-of-rotation of the gyro motion from the magnetic field $\bhat$ to the guiding-centre direction of motion $\bb{P}/P$. Hence, to first order in guiding-centre theory, synchrotron radiation is emitted in an approximately circular cone, with opening angle equal to the pitch angle $\thetap=\arctan(\pperp/\ppar)$, around the guiding-centre momentum vector $\bb{P}$.

\subsection{Implementation and validation}
The conclusions from the analysis of the electron momentum vector suggest that the cone model, implemented previously \SOFT~\citep{Hoppe2018a}, can be modified to take guiding-centre drifts into account by including the first-order drift terms in the \emph{guiding-centre} momentum vector $\bb{P}$, and consider the radiation to be emitted in an infinitesimally thin circular cone around $\bb{P}$. The method is however not exact since the $\mathcal{O}(\epsilon)$ terms in~\eqref{eq:momvec} can break gyrotropy, making the emission cone non-circular. Therefore, when we implement the modified cone model in \SOFT, we also implement a hybrid full-orbit/guiding-centre model that can indicate when the cone model assumptions are violated.

\SOFT\ computes the radiation observed from an ensemble of particles with distribution function $f_{\rm gc}(r,\bb{p})$ by evaluating the integral
\begin{equation}\label{eq:softint}
    P = \int\Theta\left(\frac{\bb{\rcp}}{\rcp}\right) \frac{\hat{\bb{n}}\cdot\bb{\rcp}}{\rcp^3} \frac{\dd P(\bb{x},\bb{p},\bb{\rcp})}{\dd\Omega} f_{\rm gc}(r,p,\mu)\,J\,\dd r \dd\tau \dd\phi\,\dd\bb{p}^{(0)}\,\dd A.
\end{equation}
Here, $\bb{\rcp} = \bb{x}_0-\bb{x}$ is the vector extending from the detector at $\bb{x}_0$ to the particle at $\bb{x}=\bb{x}(r,\tau,\phi)$, $\bb{p}^{(0)}$ is the particle momentum at the outer mid-plane, $\nhat$ is the detector surface normal, $\Theta$ is a step function that is one whenever $\bb{\rcp}/\rcp$ lies in the detector field-of-view, $\dd P/\dd\Omega$ is the angular distribution of radiation emitted by each particle, and $J$ is the Jacobian for the full coordinate transformation (including the guiding-centre transformation). The spatial coordinates used are the minor radius of the flux surface in the outer midplane, $r$; the time coordinate along a guiding-centre orbit $\tau$; and the toroidal angle $\phi$. The momentum $\bb{p}$ is therefore a function of both $\bb{x}$ and $\bb{p}^{(0)}$.

Note that although we use a guiding-centre transformation, the integrand~\eqref{eq:softint} is still dependent on the \emph{particle} position and momentum $\bb{x}$ and $\bb{p}$. This means that we have to use equation~\eqref{eq:momvec} and the corresponding expression for the Larmor radius vector $\bb{\rho}$ to consistently evaluate~\eqref{eq:softint} to $\mathcal{O}(\epsilon)$. However, since it would be prohibitively computationally expensive to introduce the drifts exactly in the cone model, we use the observation that the primary effect of the $\mathcal{O}(\epsilon)$ terms is to rotate the circular cone of radiation around the guiding-centre velocity vector, thus we simply modify the cone model according to
\begin{equation}\label{eq:conemod}
    \frac{\dd P}{\dd\Omega}\sim\delta\left( \bb{\rcp}\cdot\bhat - \rcp\cos\thetap \right)
    \quad\longrightarrow\quad
    \frac{\dd P}{\dd\Omega}\sim\delta\left( \bb{\rcp}\cdot\bb{P} - \rcp P\cos\thetap \right),
\end{equation}
where $\bhat$ denotes the local magnetic field unit vector and $\cos\thetap = \bb{p}\cdot\bhat$.

To verify that the approximation~\eqref{eq:conemod} gives sufficiently accurate result, we now also evaluate~\eqref{eq:softint} using a hybrid approach. Instead of calculating a set of guiding-centre orbits which are used to evaluate~\eqref{eq:softint} together with equation~\eqref{eq:momvec}, we calculate a set of particle orbits, from which $\bb{\rcp}$ and $\bb{p}$ are obtained exactly. While this approach may at first seem to be more accurate, we point out that equation~\eqref{eq:softint}, which is still used in this approach, was derived by making a guiding-centre transformation. Thus, the use of particle orbits to evaluate~\eqref{eq:softint} is not expected to be quantitatively accurate and, in addition, the Jacobian $J$ causes the amount of detected radiation to be weighed incorrectly. The purpose of the validation is therefore to ensure that the size, shape and position of the radiation pattern is correct, and this only requires that the guiding-centre drifts and gyro-orbit can be evaluated correctly. If the assumptions of the modified cone model~\eqref{eq:conemod} are violated, we expect the synchrotron pattern shapes calculated by the two models to deviate significantly.

\begin{figure}
    \centering
    \begin{overpic}[width=0.49\textwidth]{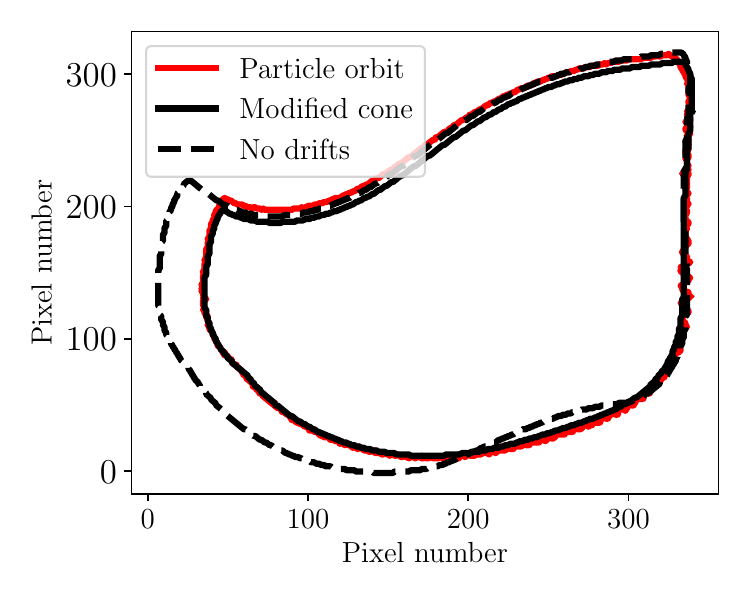}
        \put(63,16){(a) $p=30m_ec$}
    \end{overpic}
    \begin{overpic}[width=0.49\textwidth]{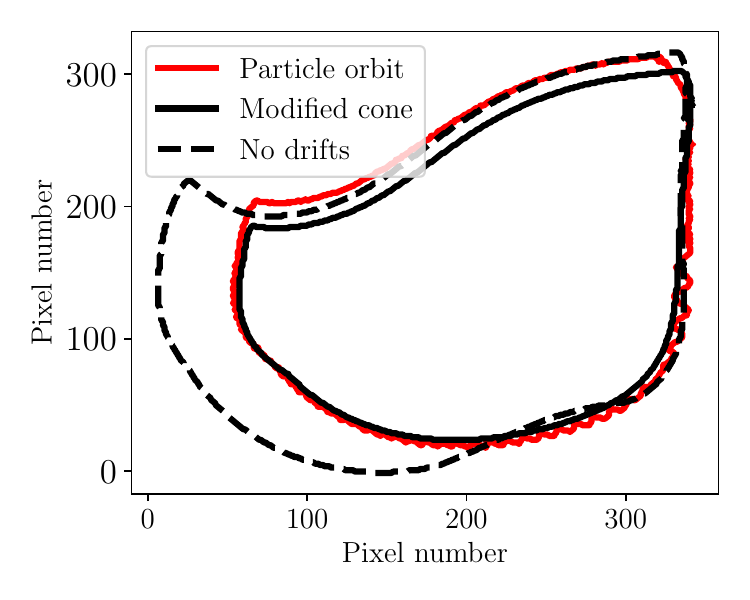}
        \put(63,16){(b) $p=60m_ec$}
    \end{overpic}
    \caption{
        Outline of the synchrotron pattern calculated using the particle orbit method (red), the modified cone model (black, solid), and the cone model neglecting guiding-centre drifts (black, dashed), at fixed pitch angle ($\theta=\SI{0.2}{rad}$) and two different momenta. In a synchrotron radiation image, every pixel enclosed by the contour would be illuminated by radiation. The main effect of the drifts is to compress the synchrotron pattern towards the low-field side, corresponding to the drift orbit shift direction.
    }
    \label{fig:conemod}
\end{figure}

Figure~\ref{fig:conemod} compares the regular cone model, the modified cone model and the particle orbit model at two different momenta and a fixed pitch angle $\thetap = \SI{0.2}{rad}$. The figures show the outline contours of a number of synchrotron images, in order to emphasize the size, shape and position of the radiation pattern. The effect of the drifts is primarily to compress the synchrotron pattern, as seen by comparing the dashed (no drifts) and solid (drifts included) curves. The modified cone model agrees mostly with the particle orbit model. As the energy is increased, the pattern resulting from the modified cone model also begins to deviate from the particle orbit model, mainly along the upper edge. This deviation is a sign of that the guiding-centre is breaking down, and gyrotropy is being violated. As an additional consequence of the theory breaking down, we observe unphysical oscillations along the lower right edge in figure~\ref{fig:conemod}b. From figure~\ref{fig:conemod} we can however conclude that the modified cone model captures the dominant effects of first order guiding-centre corrections, allowing us to more accurately model high-energy runaway scenarios without a significant increase in numerical complexity.

%% file: appendixB.tex
\section{Estimation of $\pstar$ and $\tstar$ evolution from kinetic theory}\label{app:kinetic theta}
By constraining the time evolution of the runaway pitch-angle distribution using kinetic theory, the evolution of the super-particle parameters, $\pstar$ and $\tstar$, can be inferred from the time variation of the experimentally measured total intensity---as done in figure \ref{fig:synchIntensity}. 

To obtain a simplified description of the synchrotron radiation emitted by runaways during the runaway plateau phase, we start with the expression for the emission from an electron moving in a uniform magnetic field: \citep{bekefi}
\begin{align}
\frac{\partial P}{\partial \omega} &= \frac{m_e c r_0}{\sqrt{3}\pi} \frac{\omega}{\gamma^2}\int_{\omega/\omega_c}^\infty K_{5/3}(x)\,\mathrm{d} x, \nonumber \\
\omega_c &= \frac{3}{2}\frac{eB}{m_e} \frac{\gamma^2}{\gamma_\parallel} = \frac{3}{2}\frac{eB}{m_e}\gamma\sqrt{1+(1-\xi^2)(p/m_e c)^2} \nonumber \\
\gamma_\parallel &= \frac{1}{\sqrt{1-\xi^2(v/c)^2}} = \frac{\gamma}{\sqrt{1+(1-\xi^2)(p/m_e c)^2}} \nonumber \\
\omega &= 2\pi c/\lambda.
\end{align}
When $p_\perp \gg m_e c$, the critical frequency---where the emission is the strongest---is approximately given by $\omega_c \approx 3eB\gamma^2\theta/(2m_e)$. Considering an observed wavelength of $\lambda \approx 700\,$nm, a representative pitch angle $\theta \approx 0.2$, and energy $\gamma \approx 50$, we find that $\omega/\omega_c \approx 8$. Then, the emission is well described by the formula evaluated at the  $\omega/\omega_c \gg 1$ limit
\begin{align}
\frac{\partial P}{\partial \omega} \approx \frac{m_e c r_0}{\sqrt{3}\pi}\frac{\sqrt{\omega_c\omega}}{\gamma^2}e^{-\omega/\omega_c}.
\end{align}
The total emission from a distribution of runaways, which we assume can be approximated as $f_\mathrm{RE} \approx F(t,\,p)\exp(-C(t,\,p)\theta^2/2)$, is given by the phase-space integral of
\begin{align}
\mathcal{P} = \exp\left( -\frac{2\omega m_e}{3eB}\frac{1}{\gamma^2\theta} - \frac{C}{2}\theta^2\right),
\end{align}
up to a factor depending on plasma and runaway parameters at most polynomially (in contrast to the dominant, exponential dependence captured by $\mathcal{P}$). At a given $\pstar$ (the precise value of which depends on the energy distribution of runaways), this function is maximized by the pitch angle $\tstar$, and takes the maximum value $\mathcal{P}(\tstar)$, given by
\begin{align}
\tstar &= \left(\frac{2\omega m_e}{3eB C\gamma^{\star2}}\right)^{1/3}, \nonumber \\
\mathcal{P}(\tstar) &= \exp\left[ -\left( \frac{\omega m_e}{3\sqrt{2}eB}\right)^{2/3}\frac{C^{1/3}}{\gamma^{\star 4/3}}\right].
\label{eq:POfThetaStar}
\end{align}

\subsection{Determination of $C$ from kinetic dynamics}
At the end of the current quench, when the plasma current is turning into its plateau phase, the electric field approaches a critical value where the runaway population is marginally sustained~\citep{breizman2014}. In that case, the momentum flux in phase space becomes negligible, and the subsequent evolution throughout the plateau phase is mainly pitch angle relaxation, described by the equation
\begin{align}
\frac{\partial f}{\partial t} = \frac{\partial}{\partial \xi}\left[(1-\xi^2)\left(-\frac{E}{p}f + \frac{\nu_D(p)}{2}\frac{\partial f}{\partial \xi}\right)\right],
\end{align}
which is similar to the equation studied by~\cite{Aleynikov2015} to determine the equilibrium distribution.
Taking integral moments in $\xi$ of this equation yields
\begin{align}
\langle 1-\xi \rangle &= \frac{\int_{-1}^1\mathrm{d}\xi \, (1-\xi) f }{\int_{-1}^1\mathrm{d}\xi \, f }, \nonumber \\
\frac{\partial \langle 1-\xi \rangle}{\partial t} &= 
\frac{\int_{-1}^1 \mathrm{d} \xi \, \left[ -(1-\xi^2) \frac{E}{p}f + \nu_D \xi f\right]}{ \int_{-1}^1\mathrm{d}\xi \, f} \nonumber \\
&\approx - \frac{2E}{p}\langle 1-\xi\rangle + \nu_D,
\end{align}
where the last relation is valid for small pitch angles $1-\xi \ll 1$. Since the initial pitch angle distribution is highly anisotropic---having been generated by an acceleration in a strong electric field---the first term will initially be negligible, $\nu_D \gg 2E\langle 1-\xi\rangle/p$. For a distribution of the form $f\propto \exp(-C\theta^2/2)$, one finds $\langle 1 - \xi \rangle = 1/C$, which then yields the time evolution
\begin{align}
C(t) = \frac{1}{\frac{1}{C(t_0)} + (t-t_0)\nu_D}\approx \frac{p^{\star 2}}{8n_\mathrm{Ar,20}(t-t_0)_\mathrm{ms}},
\end{align}
In a cold post-disruption plasma where the pitch-angle scattering of electrons is dominated by collisions with argon impurities, the  diffusion rate in pitch-angle is given by~\cite{Hesslow2018generalized}:
\begin{align}
\nu_D &\approx 4\pi n_\mathrm{Ar} c r_0^2\frac{\gamma}{p^3} \ln 90p \nonumber \\
&\approx \frac{8n_\mathrm{Ar,20}}{p^{\star2}}\,\times (1 \,\text{ms}^{-1}),
\end{align}
where the argon density is expressed in units of $10^{20}$m$^{-3}$ in the last expression.

Combining this result with equation~\eqref{eq:POfThetaStar}, we finally obtain the following expression for the time evolution of the total emitted synchrotron power:
\begin{equation}
\begin{aligned}
\mathcal{P} &=  \exp\left[ -\left( \frac{\omega m_e}{3\sqrt{2}eB}\right)^{2/3}\frac{1}{\gamma^{\star 4/3}}\frac{1}{\left[\frac{1}{C(t_0)} + (t-t_0)\nu_D \right]^{1/3}}\right]. \\
\end{aligned}
\end{equation}